\documentclass[twocolumn,showpacs,preprintnumbers,amsmath,amssymb]{revtex4}

\usepackage{graphicx,epsf}
\usepackage{dcolumn}
\usepackage{bm}

\begin{document}


\title{Optical phonon scattering and theory of magneto-polarons \\
in a quantum cascade laser in a strong magnetic field}

\author{Yu  Chen$^{1,2}$, N. Regnault$^1$, R. Ferreira$^1$, Bang-Fen Zhu$^2$ and G. Bastard$^1$}
\affiliation{$^1$Laboratoire Pierre Aigrain, Ecole Normale Sup\'erieure, CNRS, 24 Rue Lhomond, F-75005 Paris, France\\
$ ^2$Department of Physics, Tsinghua University, Beijing 100084, China}

\date{\today}

\begin{abstract}
We report a theoretical study of the carrier relaxation in a quantum
cascade laser (QCL) subjected to a strong magnetic field.
Both the alloy (GaInAs) disorder effects and the Fr\"ohlich interaction
are taken into account when the electron energy
differences are tuned to the longitudinal optical (LO) phonon energy.
In the weak electron-phonon coupling regime, a Fermi's golden rule computation
of LO phonon scattering rates shows a very fast non-radiative relaxation
channel for the alloy broadened Landau levels (LL's).
In the strong electron-phonon coupling regime, we use a magneto-polaron
formalism and compute the electron survival probabilities
in the upper LL's with including increasing numbers of LO phonon modes
for a large number of alloy disorder configurations.
Our results predict a nonexponential decay of the upper level
population once electrons are injected in this state.
\end{abstract}


\maketitle

\section{Introduction}
A quantum cascade laser (QCL) is a semiconductor laser
with specially designed cascade quantum structure to realize unipolar
carrier transport and intersubband optical transition, and thus to
achieve long-wavelength lasing by overcoming
the bandgap limit of the material~\cite{QCL}. Since the first
observation of the population inversion in a QCL~\cite{Faist}, the mid-infrared,
far-infrared and even THz QCL's have been realized though few of
them operate at room temperature~\cite{Williams}. It has been
observed that the performances of the QCL's deteriorate with
temperature quickly when the emission wave length gets longer, a
signature of an increasingly detrimental thermal activation of
non-radiative losses. Therefore it is important to
understand and control the non-radiative paths from the upper
state of the lasing transition. A magnetic field ${B}$ applied parallel
to the growth axis has been used as an external parameter to monitor the
characteristic of a QCL. In particular, it modulates the output power of
the laser with an oscillation period proportional
to ${1/B}$. This output is sensitive to both
elastic and inelastic scatterings, which introduce non-radiative
relaxation channels to carriers in the Landau level
(LL)~\cite{Ulich,Blaser,Becker, Smirnov,Scalari,Pere}. Thus it is of
interest to study theoretically the relaxation mechanisms of the QCL
in the presence of a quantizing magnetic field.
Calculations of the LO phonon emission rates in these disordered
systems have rarely been reported~\cite{Brown2000,Leuliet,Ivana}.
 Yet, in the ${\rm Ga_{0.47}In_{0.53}As}$ active-region material
 the alloy effect nearly reaches its maximum value, with broadening
 of the LL up to several $meV$s. Therefore, it is worth devoting more
 efforts to the understanding whether the notion of phonon emission
 remains meaningful in LL quantized material.

Assuming a LL structure at high magnetic field, the problem could be
tackled in two radically different ways. In the weak electron-phonon
coupling case, the electron states are initially computed by including
alloy scattering potentials and the phonon emission rates are then
calculated using the Fermi's golden rule. This traditional perturbative
computation would give an estimate of the exponential decay rate of
the upper level electron population. On the contrary, LL electrons
and LO phonons form magneto-polaron states when their coupling is
strong enough. Similar effects appear recurrently in the optical properties of
quantum dots (QD's), ${\it e.g.}$~\cite{Fomin,Grange}. In this mixed mode
description the electron-LO-phonon interaction is initially
taken into account exactly. Hence, the very notion of
phonon scattering/emission becomes irrelevant. We shall examine the
magneto-polaron case where the alloy scattering broadens the
magneto-polaron states and compute the time-dependent survival
probabilities in the upper LL's for an electron initially
in one of the alloy broadened upper LL's. This electron-phonon strong coupling
theory leads to a distinguishably different prediction for the
carrier relaxation process as composed to the ordinary phonon emission picture,
and predicts a nonexponential population decay from the upper levels.

\section{LO phonon emission of the alloy broadened LL's}

For mid-infrared QCL's, it is a reasonable assumption that the alloy
scattering destroys neither the subband structure at zero field nor
the LL structure at high magnetic fields. So we consider only a few
subbands and their LL's. Leuliet {\it et al}
demonstrated that the alloy
scattering effect in the barrier is negligible as compared to that in
the well~\cite{Leuliet}. We correspondingly model the active region of
 a QCL as a single quantum well (QW) clad between infinite potential barriers
along the $z$ direction. A uniform
magnetic field applied parallel to the quantum confinement direction
is considered in Landau gauge (with the vector potential ${\vec A=Bx\hat{y}}$). Moreover, we neglect the spin Zeeman
effect. Thus, in the effective mass model the unperturbed degenerate LL eigenstates of
the ${l}$th subband read

\begin{eqnarray}
\big<\vec
r\big|E_l,n,k_y\big>&=&\frac{1}{\sqrt{L_y}}e^{ik_yy}\chi_l(z)\varphi_n(x+\lambda^2k_y),{}
\nonumber\\
{}\varepsilon_{l,n}&=&E_{l}+\Big(n+\frac{1}{2}\Big)\hbar\omega_{c},
\end{eqnarray}
where ${\chi_l}$ as the
wavefunction of the ${l}$th subband,
${\varphi_n}$ the ${n}$th Hermite function, ${\lambda=\sqrt{{\hbar}/{eB}}}$ the magnetic
length and ${\omega_{c}={eB}/{m^{*}}}$ the cyclotron frequency.
We take ${m^{*}=0.05m_{0}}$ for electron effective mass and ${L_z=12.33nm}$
for quantum well width, which gives the zero-field subband separation
${E_2-E_1=147.5meV}$. The LL degeneracy (per spin) is
${D=L_xL_y/{(2\pi\lambda^{2})}}$.

\begin{figure}
\includegraphics[height=.33\textheight, angle=270]{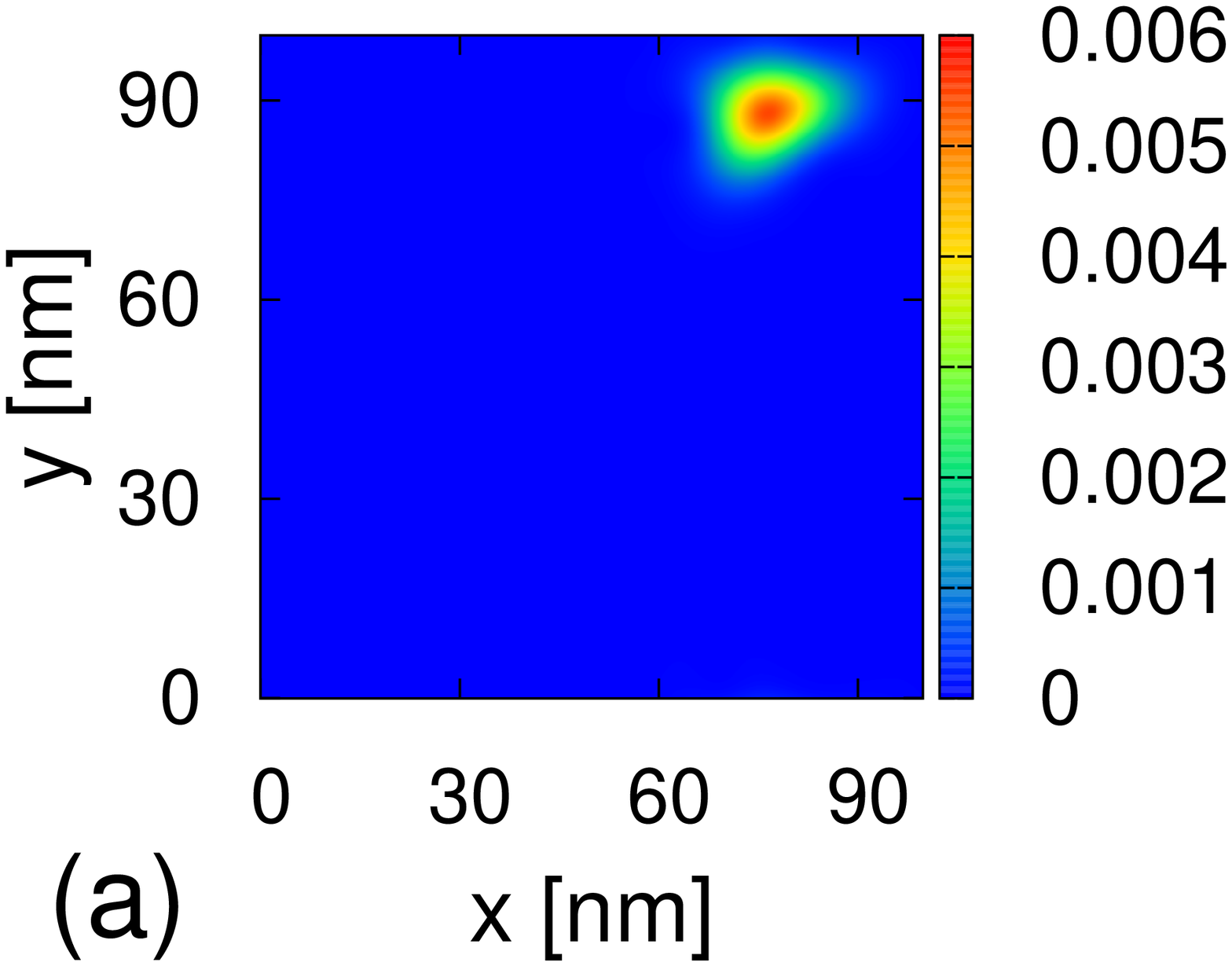}
\includegraphics[height=.33\textheight, angle=270]{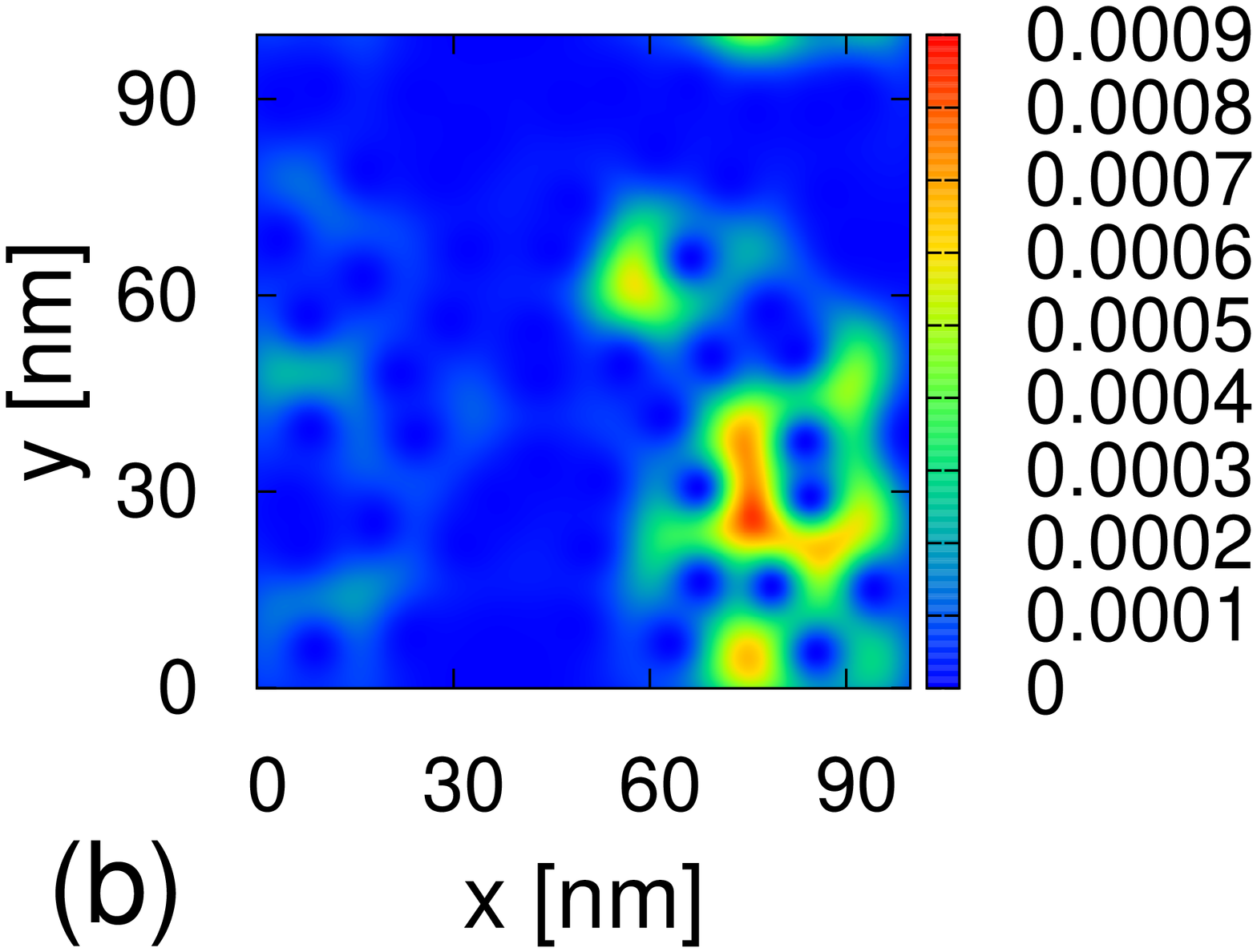}
\includegraphics[height=.33\textheight, angle=270]{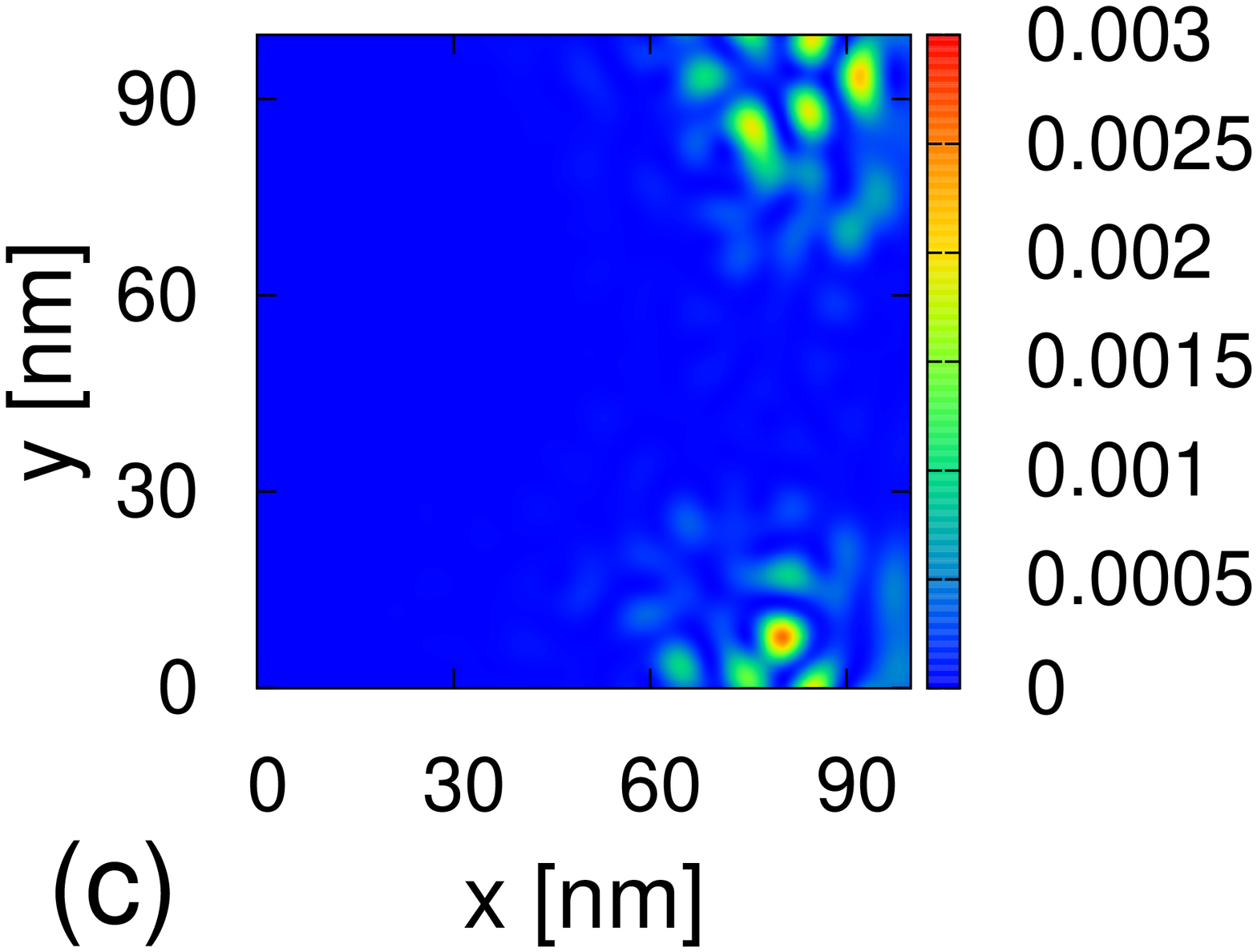}
\includegraphics[height=.33\textheight, angle=270]{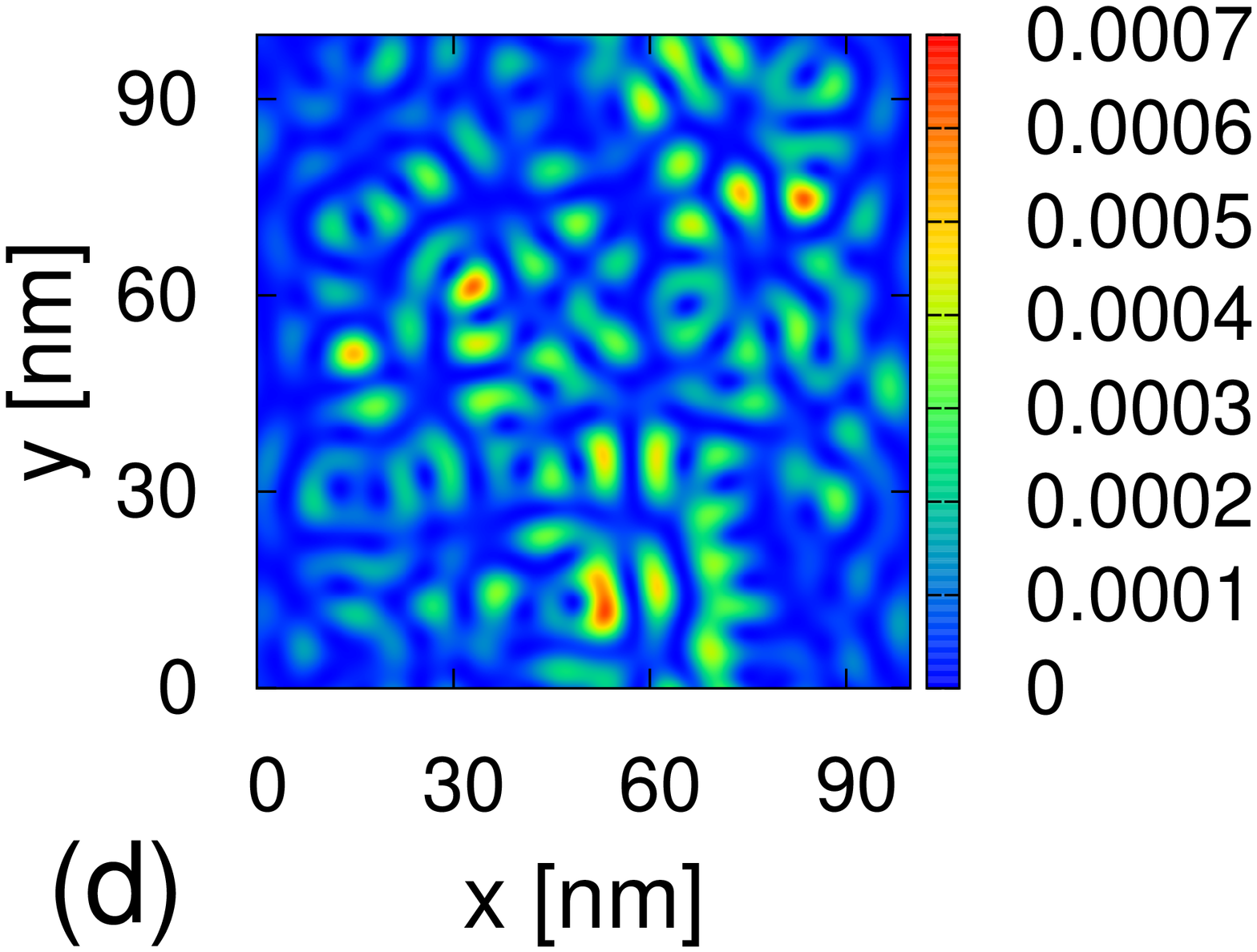}
  \caption{(Color online) Typical in-plane electron density distributions (in \AA${^{-2}}$) for four states in the set of independently broadened
  LL's at 24.58T: (a) DOS tail of ${\big|E_2,0\big>}$; (b) DOS max of ${\big|E_2,0\big>}$; (c) DOS tail of ${\big|E_1,2\big>}$; (d) DOS max of ${\big|E_1,2\big>}$.}
\end{figure}

For a numerical calculation of the alloy scattering effect in the
${\rm Ga_{1-x}In_xAs}$ well, we partition the system (a large box of
${\Omega_0=L_x\times L_y\times
L_z=99.79nm\times99.79nm\times12.33nm}$) into tiny unit cells
(${\omega_{0}=2.93}$\AA${\times2.93}$\AA${\times5.87}$\AA)~\cite{BrownAustin:2000}.
In each cell the scattering potential is a random variable which
equals to x${\Delta V}$ with probability ${1-}$x and to
${-(1-}$x${)\Delta V}$ with probability x, where x${=0.53}$ for
${\rm Ga_{0.47}In_{0.53}As}$ and ${\Delta V=0.6eV}$~\cite{Jang}. On
the scale of LL's or $z$ dependent wavefunction extensions, the
alloy fluctuations act like delta scatterers and we assume there is
no correlation between the x values of different cells. Thus the
alloy potential is given by~\cite{Wave}
\begin{equation}
V_{alloy}=\Delta V\omega_{0}\Big(\sum_{{\vec R_{\rm{Ga}}}}{\rm{x}}\delta (\vec
r-\vec R_{\rm{Ga}})-\sum_{\substack{\vec R_{\rm{In}}}}(1-{\rm{x}})\delta (\vec r-\vec R_{\rm{I
n}})\Big).
\end{equation}
With a given configuration of the alloy disorder, we calculate the
alloy broadened LL's by diagonalizing the alloy potential in the
basis of the degenerate LL's (for details, see
Ref.~\cite{BrownAustin:2000}). With our sample box this means
${[D]=59}$ states per LL at 24.58T. The diagonalization has been
done with ${\rm{N}=100}$ samples with randomly distributed alloy
atoms.

\begin{figure}
\includegraphics[height=.335\textheight, angle=270]{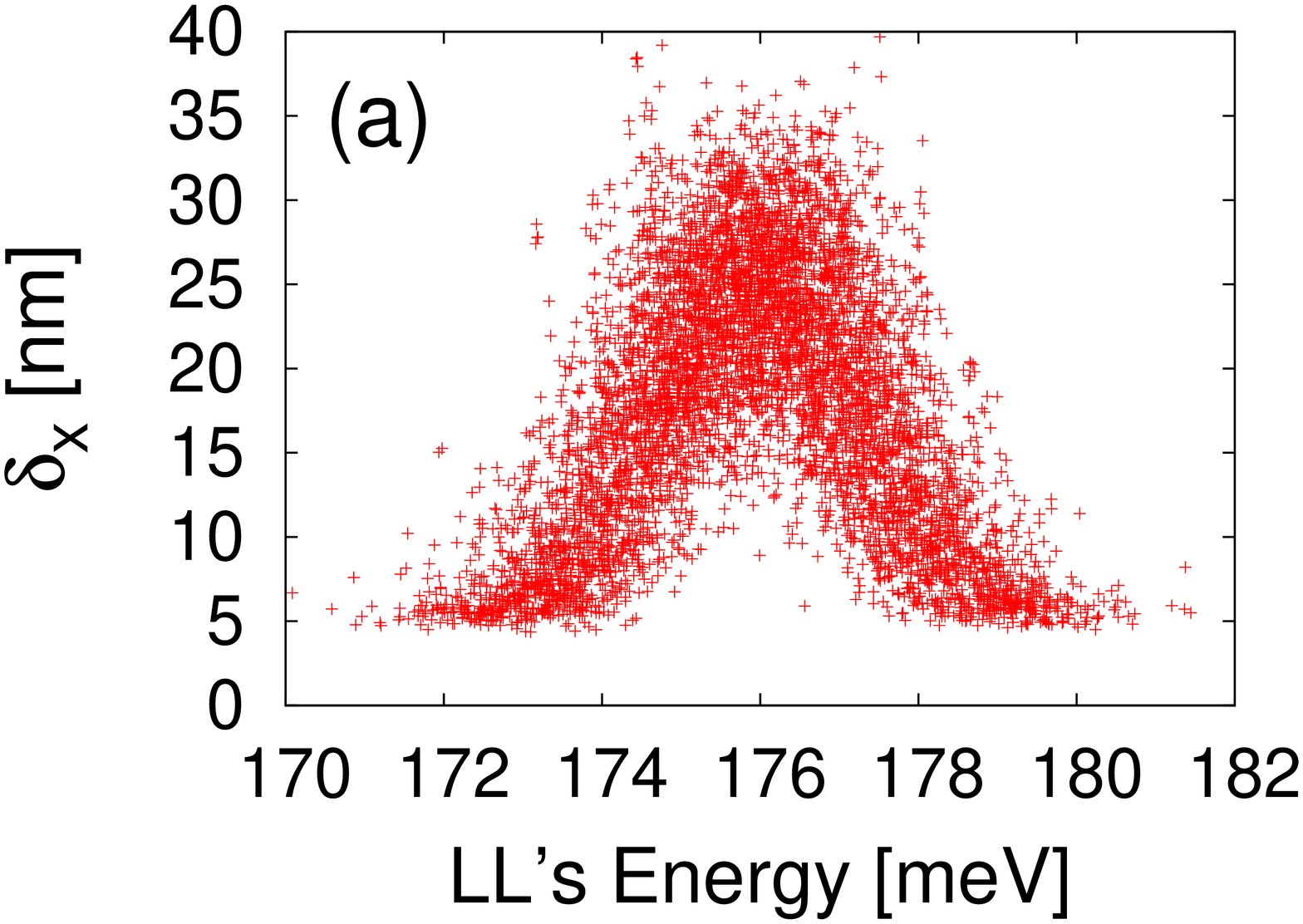}
\includegraphics[height=.335\textheight, angle=270]{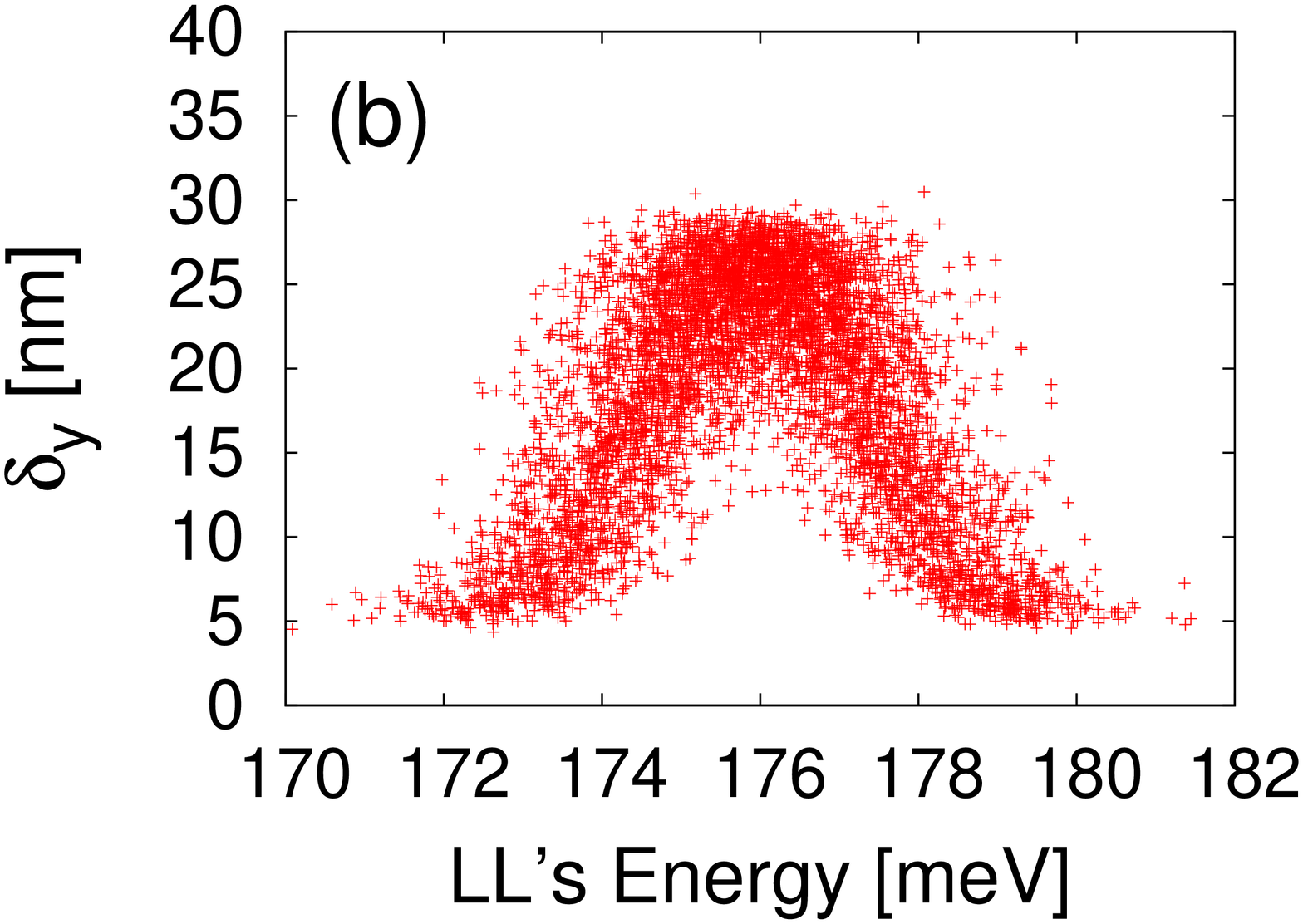}
\includegraphics[height=.335\textheight, angle=270]{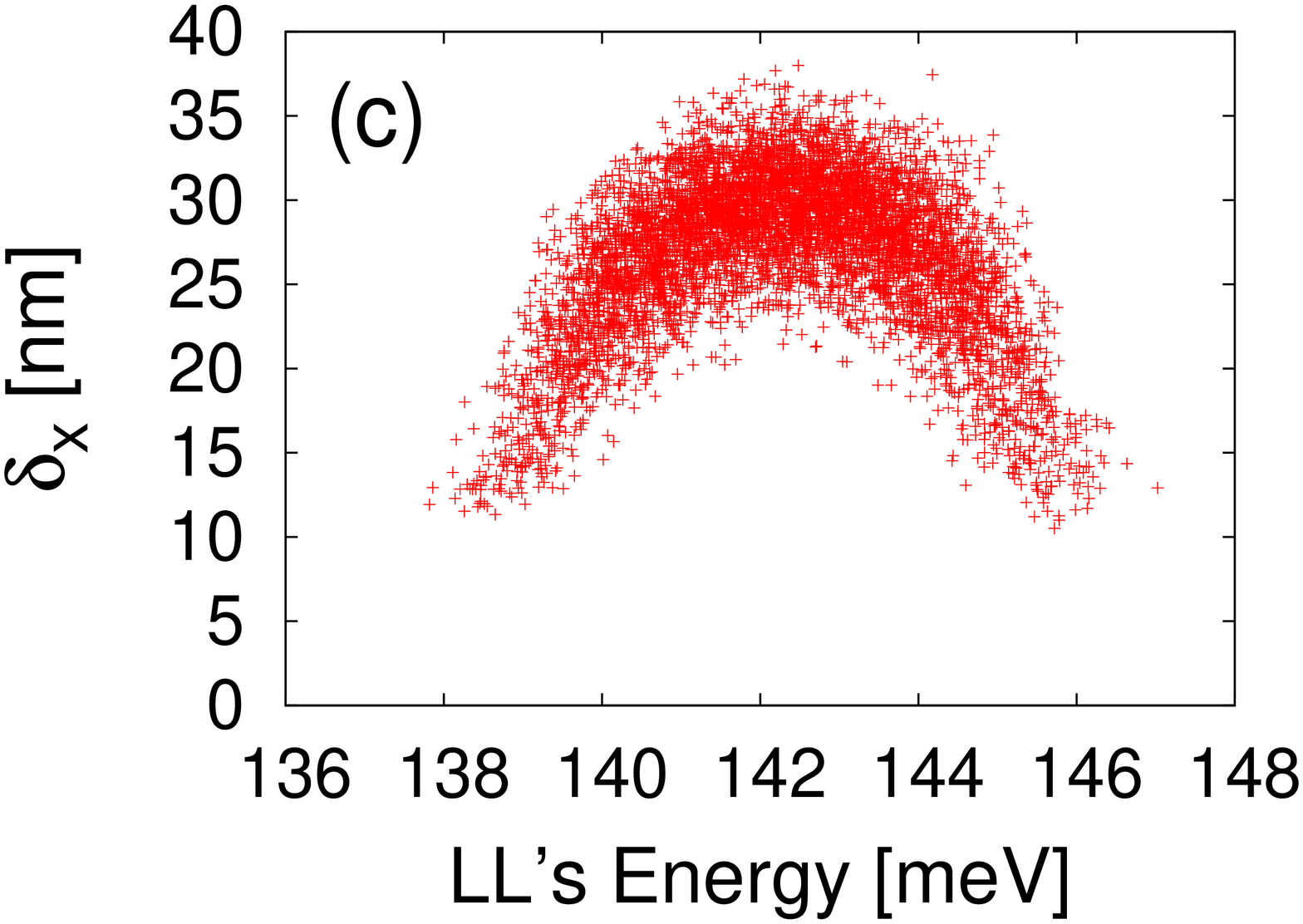}
\includegraphics[height=.335\textheight, angle=270]{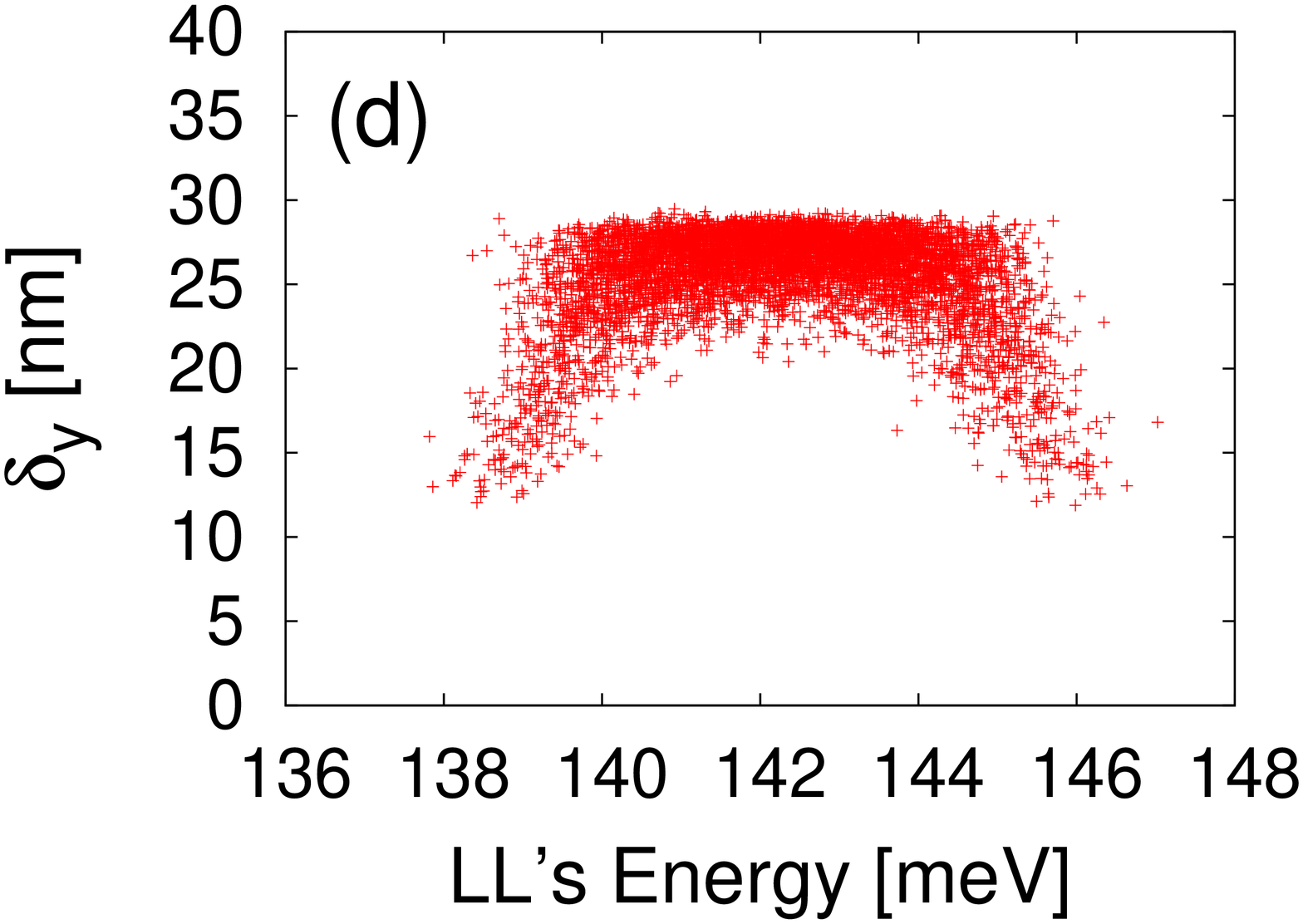}
  \caption{(Color online) Position uncertainties of the independently alloy broadened LL's at 24.58T: (a) ${\delta_x}$ of ${\big|E_2,0\big>}$; (b) ${\delta_y}$ of ${\big|E_2,0\big>}$; (c) ${\delta_x}$ of ${\big|E_1,2\big>}$ ; (d) ${\delta_y}$ of ${\big|E_1,2\big>}$.}
\end{figure}

The numerical results (Figs. 1 and 2) demonstrate the validity of the
LL structure and clearly show the localization of the electrons at the
broadened LL's tails as well as the extended behavior at the maximum of
the density of states (DOS). 
Fig. 1 shows the in-plane electron density distributions ${P(\vec
\rho)}$ for four states of one particular alloy configuration, as
defined by
\begin{equation}
P_{l,n,\nu}(\vec \rho)=\int \chi_l^2 (z)\mathrm{d}z\big|\psi_{n,\nu} (\vec \rho) \big|^2,
\end{equation}
where ${\big<\vec r\big|E_l,n,\nu\big>=\chi_l(z)\psi_{n,\nu}(\vec \rho)}$ is
the ${\nu}$th disordered state related to ${\big|E_{l},n\big>}$ (a short notation
for all unperturbed or disordered states related to the ${l}$th subband and ${n}$th LL's).
For each of the ${\big|E_l,n,\nu\big>}$ states of a particular run,
we have computed the ${x}$- and ${y}$-uncertainties,
\begin{eqnarray}
\delta_{x}&=&\sqrt{\big<x^2\big>-\big<x\big>^2},
\\
\delta_{y}&=&\sqrt{\big<y^2\big>-\big<y\big>^2}.
\end{eqnarray}
Fig. 2 shows ${\delta_{x}}$ and ${\delta_y}$ for all the
independently alloy disordered states related to ${\big|E_2,0\big>}$
or ${\big|E_1,2\big>}$ at 24.58T (${[D]=59}$; ${\rm{N}=100}$
disorder configurations). For an unbroadened LL
${\big|E_l,n,k_y\big>}$ with the electron entirely delocalized along
the ${y}$ axis over a segment ${L}$ there is
${\delta_{y}=\frac{L}{2\sqrt{3}}}$ and
${\delta_x^{(n)}=\lambda\sqrt{n+\frac{1}{2}}}$. For ${L=10^2 nm}$
and ${B=24.58T}$, this means ${\delta_y=28.8nm}$,
${\delta_x^{(0)}=3.66nm}$ and ${\delta_x^{(2)}=8.18nm}$. Fig. 2
shows that the tail states are considerably more localized than the
central states. Indeed, for the latter ${\delta_y}$ is close to the
expected value for a plane wave. We also note that
${\delta_x^{(0)}}$ and ${\delta_x^{(2)}}$ are considerably larger
than ${3.66nm}$ and ${8.18nm}$, respectively, and in fact close to
${\delta_y}$. This proves that the extended states of the alloy
broadened LL's are very different from the unperturbed
${\big|E_l,n,k_y\big>}$ states but display similar extensions along
the ${x}$ and ${y}$ directions (the small discrepancies are due to
the different boundary conditions along the ${x}$ and ${y}$
directions).

We consider the electron-phonon interaction here only due to the
GaAs-like LO phonons, because in the ${\rm Ga_{0.47}In_{0.53}As}$
alloy the Fr\"ohlich interaction is dominant and the potential of
GaAs-like modes dominates over that of the InAs-like
modes~\cite{Wang}. We take the zone-center LO mode frequency as
${\hbar\omega_{LO}=33.7 meV}$ and the Fr\"ohlich interaction as
\begin{equation}
H_{e-ph}=\sum_{\vec q}\Big(i\frac{g}{q}e^{-i\vec q\cdot\mbox{}\vec
r}b_{\vec q}^{+}+h.c.\Big)
\end{equation}
where ${b_{\vec q}^{+}}$ is the creation operator of an LO phonon with
 a wavevector of ${\vec q}$ and the Fr\"ohlich factor is
${g=\sqrt{\frac{e^2\hbar\omega_{LO}}{2\varepsilon_{0}V_{cr}}(\varepsilon_{\infty}^{-1}-\varepsilon_{s}^{-1})}}$
with ${\varepsilon_{\infty}=11.6}$ and ${\varepsilon_{s}=13.3}$ as
the high-frequency and static relative dielectric constants of GaAs
material, respectively, and ${V_{cr}}$ the crystal volume.

\begin{figure}
  \includegraphics[height=.34\textheight,angle=270]{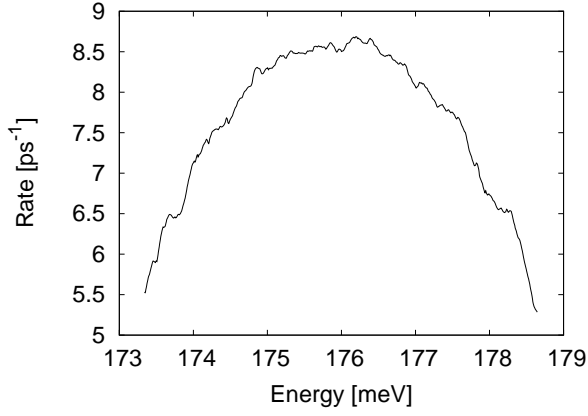}
  \caption{LO phonon emission rate from alloy broadened LL's ${\big|E_2,0\big>}$ to alloy broadened LL's ${\big|E_1,2\big>}$ at
${B_2=24.58T}$, as a function of the energy of the initial state.}
\end{figure}

In the absence of the alloy-disorder, the unperturbed factorized
states ${{\big|E_{2},n_2=0,k_{y}\big>\otimes\big|0_{LO}\big>}}$ and
${\big|E_{1},n_1=p \neq 0,k_{y}'\big>\otimes\big|1_{LO}\big>}$ cross
at the field ${B_{p}=m^{*}(E_2-E_1-\hbar\omega_{LO})/(pe\hbar)}$ for
any values of ${k_y}$ and ${k_y'}$. Let us consider the ${p=2}$
resonance, for which the turn off of QCL's lasing has been
numerically calculated under the assumption of a dominant
inhomogeneous broadening~\cite{Leuliet}. We can calculate the
scattering rate due to the electron-phonon interaction from one of
the alloy broadened ${\big|E_2,0\big>}$ LL states towards any of the
${\big|E_1,2\big>}$ broadened states assuming the Fermi's golden
rule holds. The scattering frequencies are averaged over different
alloy configurations. As shown in Fig. 3, at resonance
(${B_2=24.58T}$) the scattering rate of this irreversible departure
from the initial state reaches a maximum value when the initial
state coincides with the center of the broadened LL's; this energy
also corresponds to the largest DOS of the final states. Note that
in spite of the fact that we deal with transitions between different
LL's associated with different subbands, the scattering rate is
quite large: we find in Fig. 3 a maximum scattering rate of ${8.66
ps^{-1}}$. This relatively large value results from the increased
DOS of final states due to the Landau quantization even in the
presence of disorder. It also points out a difficulty in applying
blindly the Fermi's golden rule since the energy uncertainty
${\hbar/\tau\approx 5.70meV}$ due to LO phonon emission is
comparable to the LL width due to alloy broadening (with the full
width at ${1/e}$ DOS max as ${\Delta\approx 5.69meV}$, see
Ref.~\cite{BrownAustin:2000} and also Sec. III).

\section{Alloy broadened magneto-polaron states and the survival probability in the upper LL's}

\subsection{Unperturbed magneto-polaron states}

\begin{figure}
\includegraphics[height=.34\textheight, angle=270]{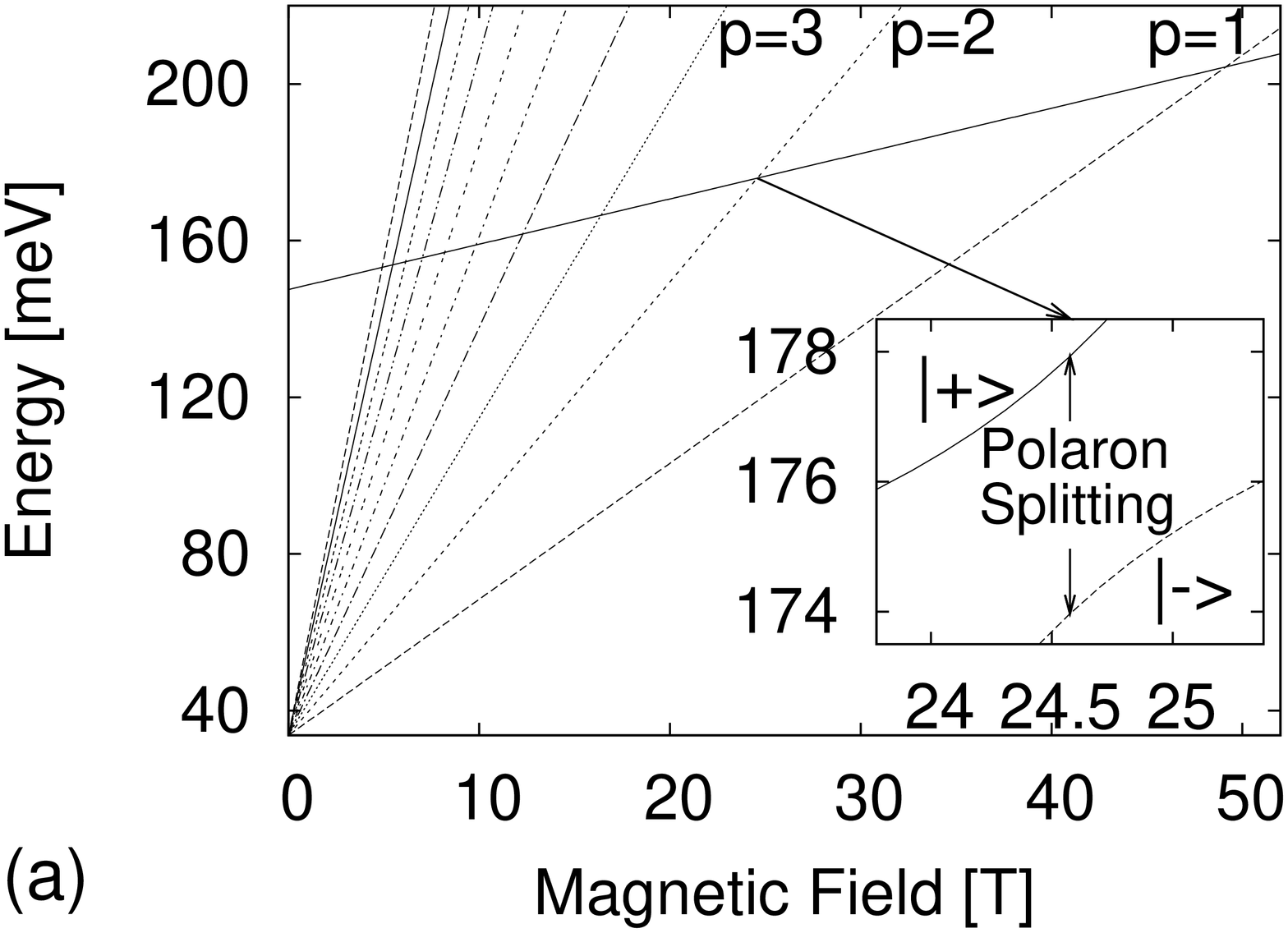}
\includegraphics[height=.34\textheight, angle=270]{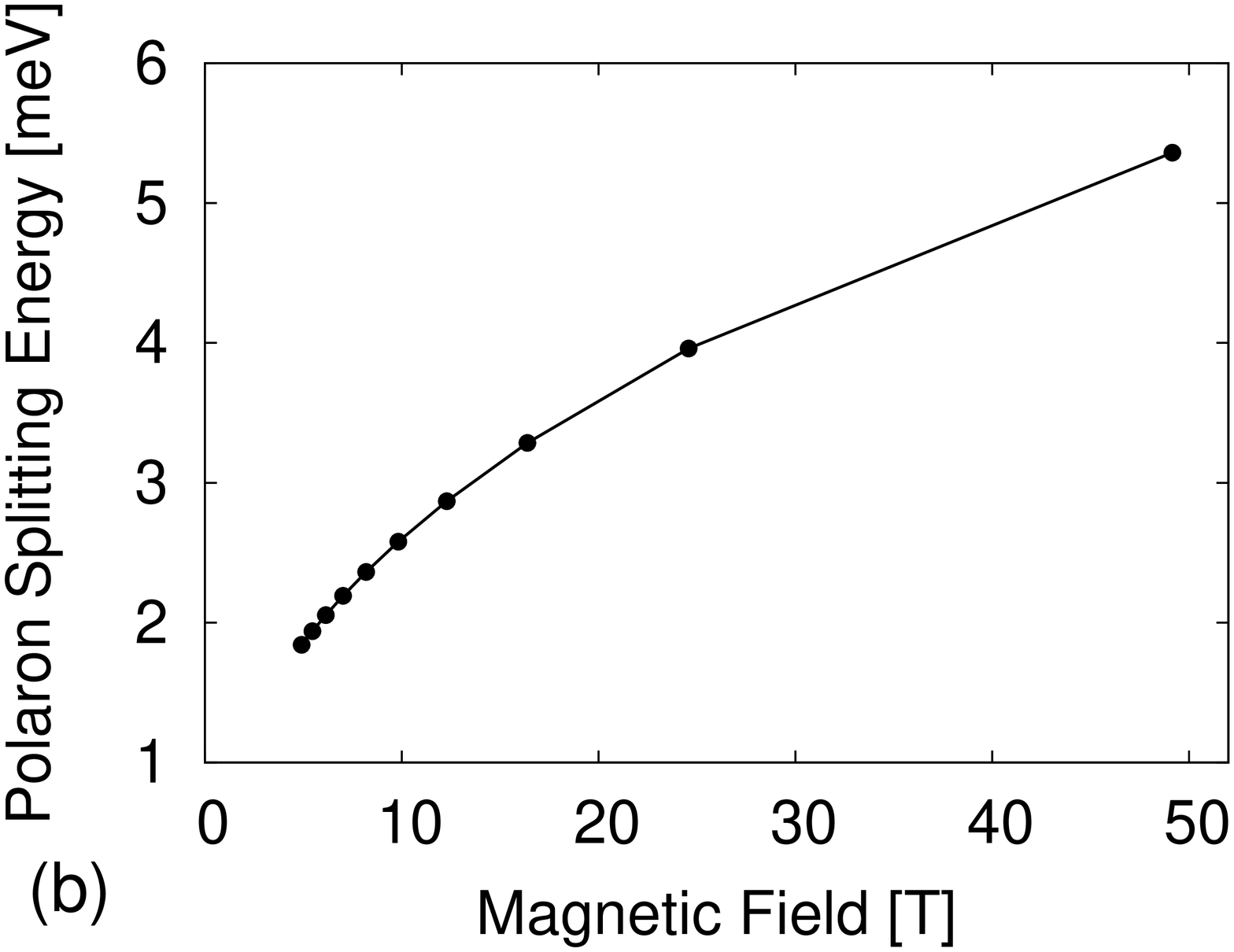}
  \caption{Illustrations of magneto-polaron states in ideal systems. (a) A schematic picture of the formation of
magneto-polaron states. Insert: energy anticrossing around ${B_2=24.58T}$.
  (b) Polaron splitting energies as a function of ${B_p}$ for ${p = 1}$ to $10$ (the line is only a guide for the eyes).}
\end{figure}

As shown in the Fig. 4, the Fr\"ohlich interaction replaces the crossing of
${{\big|E_{2},0\big>\otimes\big|0_{LO}\big>}}$ and
${\big|E_{1},p\big>\otimes\big|1_{LO}\big>}$ by an anticrossing
and leads to the formation of mixed magneto-polaron states. These
are obtained by diagonalizing the Fr\"ohlich interaction in the
factorized states manifold~\cite{Li}, and read:
\begin{eqnarray}
\big|\pm p, k_y\big>&=&\alpha_{\pm p}\big|E_2,0,k_y\big>\otimes\big|0\big>{}
\nonumber\\
{}&+&\sum_{\vec q}\beta_{\pm p,\vec q}\big|E_1,p,k_y-q_y\big>\otimes\big|1_{\vec
q}\big>,{}
\nonumber\\
{}\alpha_{\pm p}=\frac{1}{\sqrt{1+\frac{\Omega^2_p}{\omega^2_{\mp p}}}}&,&
\beta_{\pm p,\vec q}=-i\frac{g\alpha_{\pm p}}{q\hbar\omega_{\mp p}}I^{*}_{p}(\vec
q),{}
 \nonumber\\
 {}\varepsilon_{\pm p}=\varepsilon_{2,0}+\hbar\omega_{\pm p}&,&
\omega_{\pm p}=\frac{\delta_{p}}{2}\pm\sqrt{\frac{\delta^2_{p}}{4}+\Omega^2_{p}},
\end{eqnarray}
where ${\pm}$ refers to the upper ($+$) and lower ($-$) polaron states in
Figure 4(a), the detuning energy is
${\hbar\delta_{p}=\varepsilon_{1,p}+\hbar\omega_{LO}-\varepsilon_{2,0}}$,
and the half polaron splitting energy is given by
\begin{equation}
 \hbar\Omega_{p}=\sqrt{\sum_{\vec q}{g^2|I_{p}(\vec q)|^2}/{q^2}},
\end{equation}
with the electron-phonon interaction coupling matrix element as
\begin{eqnarray}
I_{p}(\vec q)&=&\big<E_2,0,k_y\big|e^{i\vec q\cdot\vec
r}\big|E_1,p,k_y-q_y\big>{}
\nonumber\\
{}&=&\frac{\big(\lambda(iq_{x}-q_{y})\big)^p}{\sqrt{2^pp!}}exp
\Big(-\frac{\lambda^2q^2_{/\!\!/}}{4}-i\lambda^2q_{x}\big(k_{y}-\frac{q_y}{2}\big)\Big){}
\nonumber\\
& &{}\times\int dz\chi^{*}_{2}(z)e^{iq_{z}z}\chi_{1}(z).
\end{eqnarray}

Actually, after the diagonalization, we find several uncoupled
${\big|E_1,p\big>}$ LL's one-phonon replica states in addition to the
magneto-polaron states. These uncoupled states 
cannot be populated by electrons entering the active region directly from the
zero-phonon states of the injection region. For simplification, in
this section we concentrate on the magneto-polaron states only (the
role of uncoupled states is discussed in Sec. IV). The
electron-phonon scattering obeys the momentum conservation law.
Since each of the unperturbed LL's has a definite ${k_y}$, the
Fr\"ohlich interaction conserves the ${k_y}$ number and, consequently,
 each magneto-polaron state has a definite ${k_y}$ value in the ideal
 system. For the case of the exact resonance between ${{\big|E_{2},0\big>\otimes\big|0_{LO}\big>}}$
 and ${\big|E_{1},p\big>\otimes\big|1_{LO}\big>}$ the upper and lower
polaron branches ${\big|\pm p,k_y\big>}$ have equal weights of
phonon components while for a finite energy detuning these weights are unequal. This
is similar to the case of strong coupling between atoms and photons
in cavity QED except that each polaron branch has a degeneracy of
the LL's~\cite{cohen}. 
The polaron splitting energy has a ${\sqrt{B}}$ dependence and
is about ${4.0meV}$ at ${B_2=24.58T}$~\cite{Li}.

\subsection{Disordered polaron DOS:\\ SCBA and numerical results}

\begin{figure}
    \includegraphics[height=.31\textheight, angle=270]{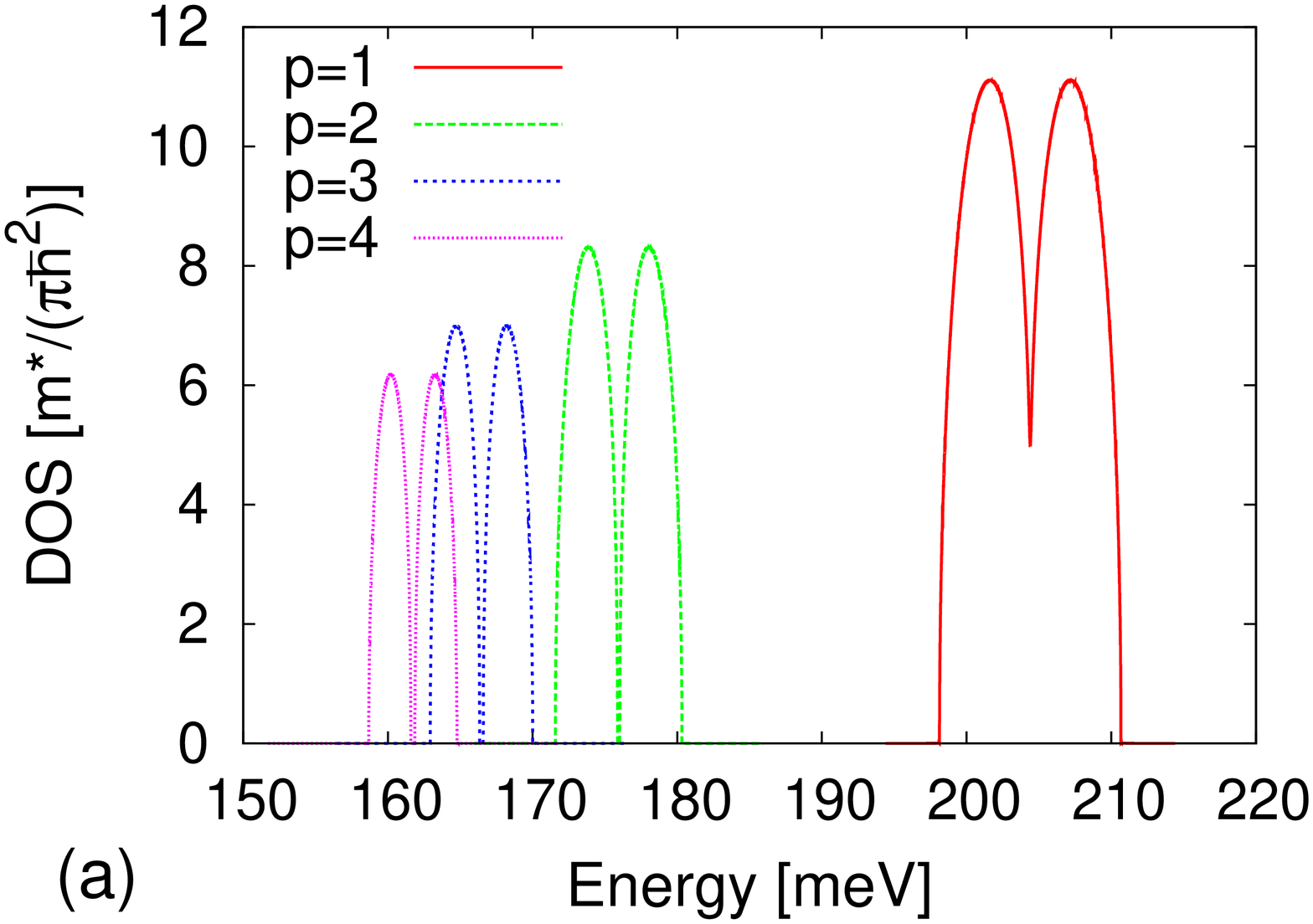}
    \includegraphics[height=.31\textheight, angle=270]{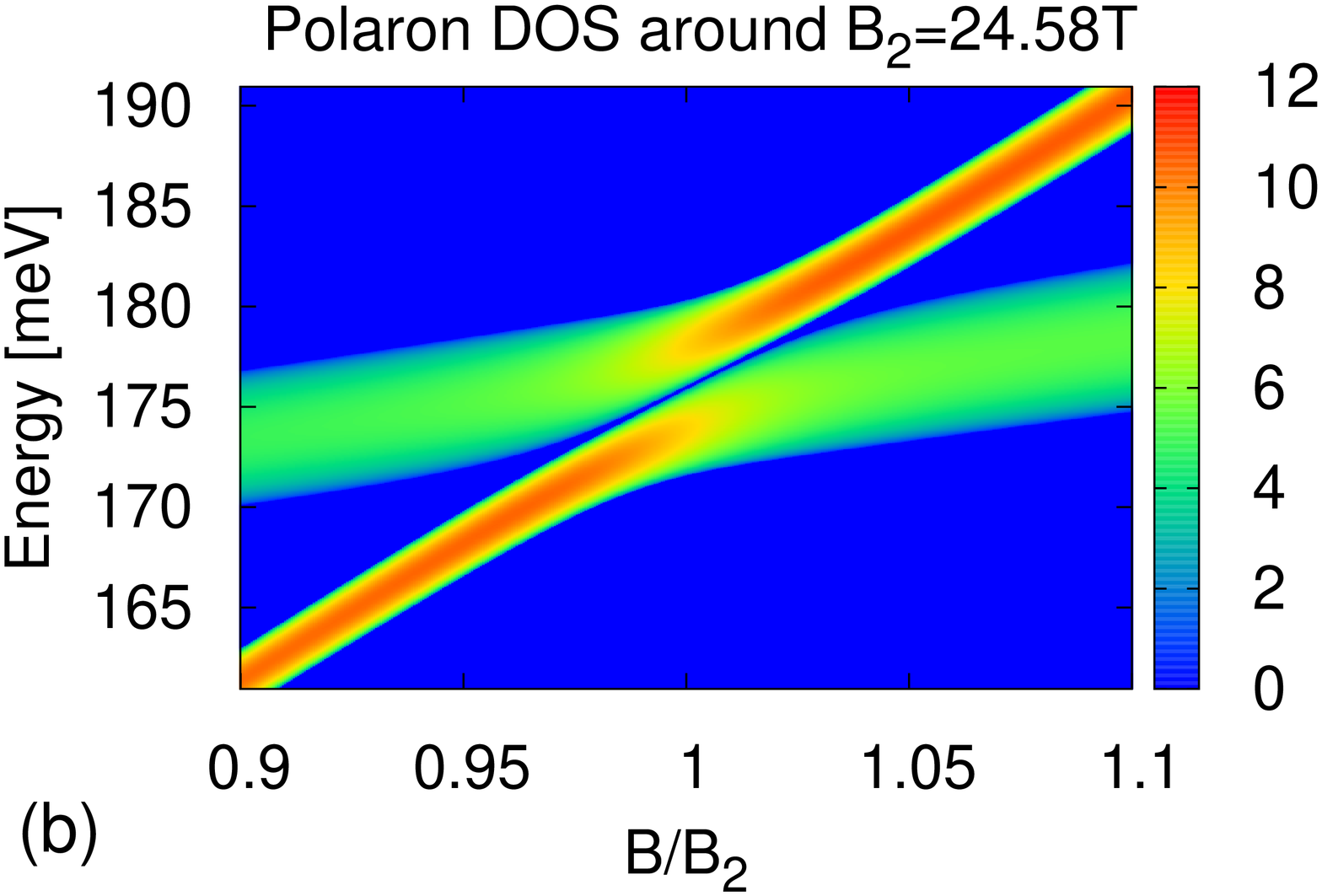}
    \includegraphics[height=.31\textheight, angle=270]{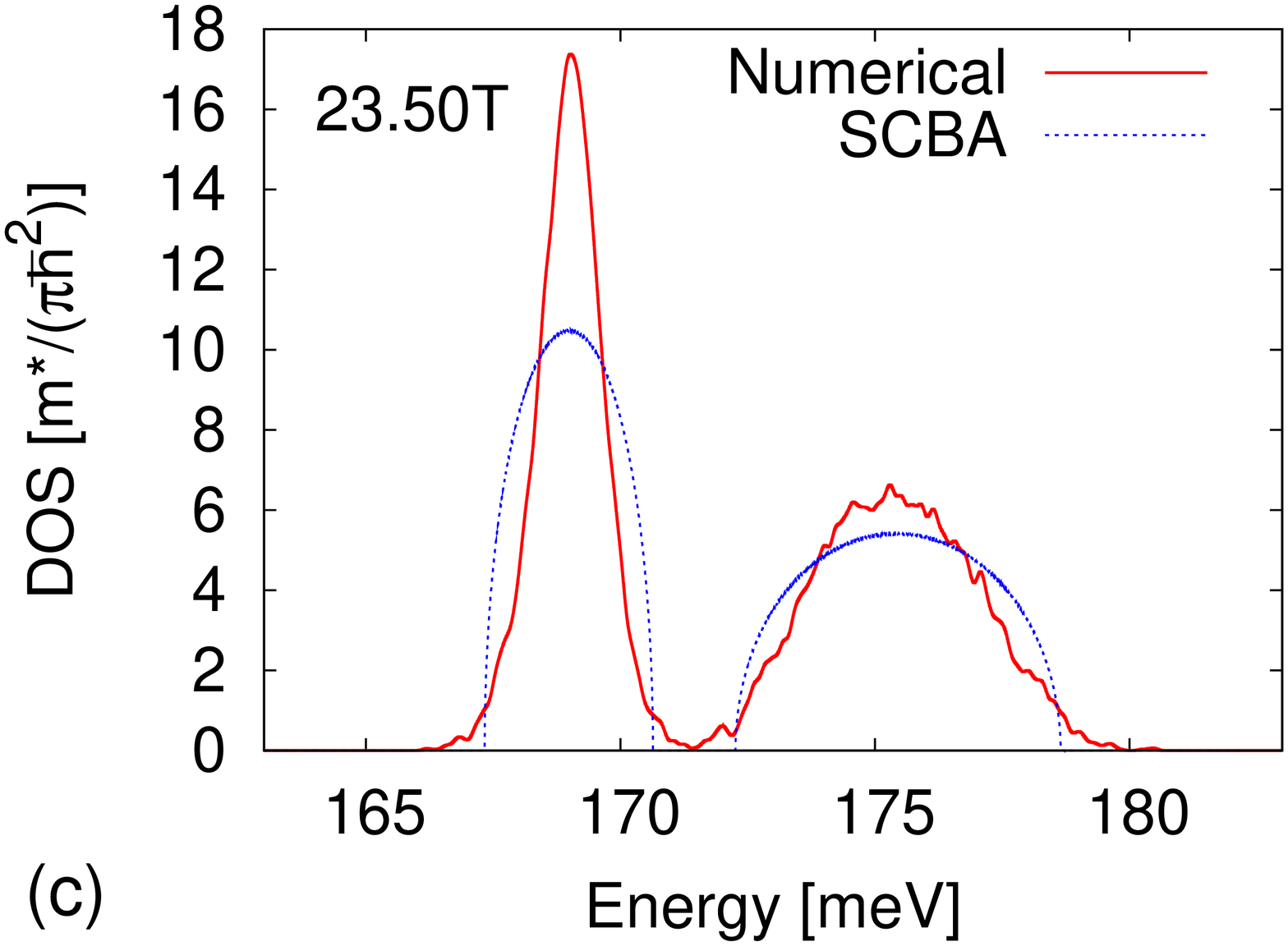}
    \includegraphics[height=.31\textheight, angle=270]{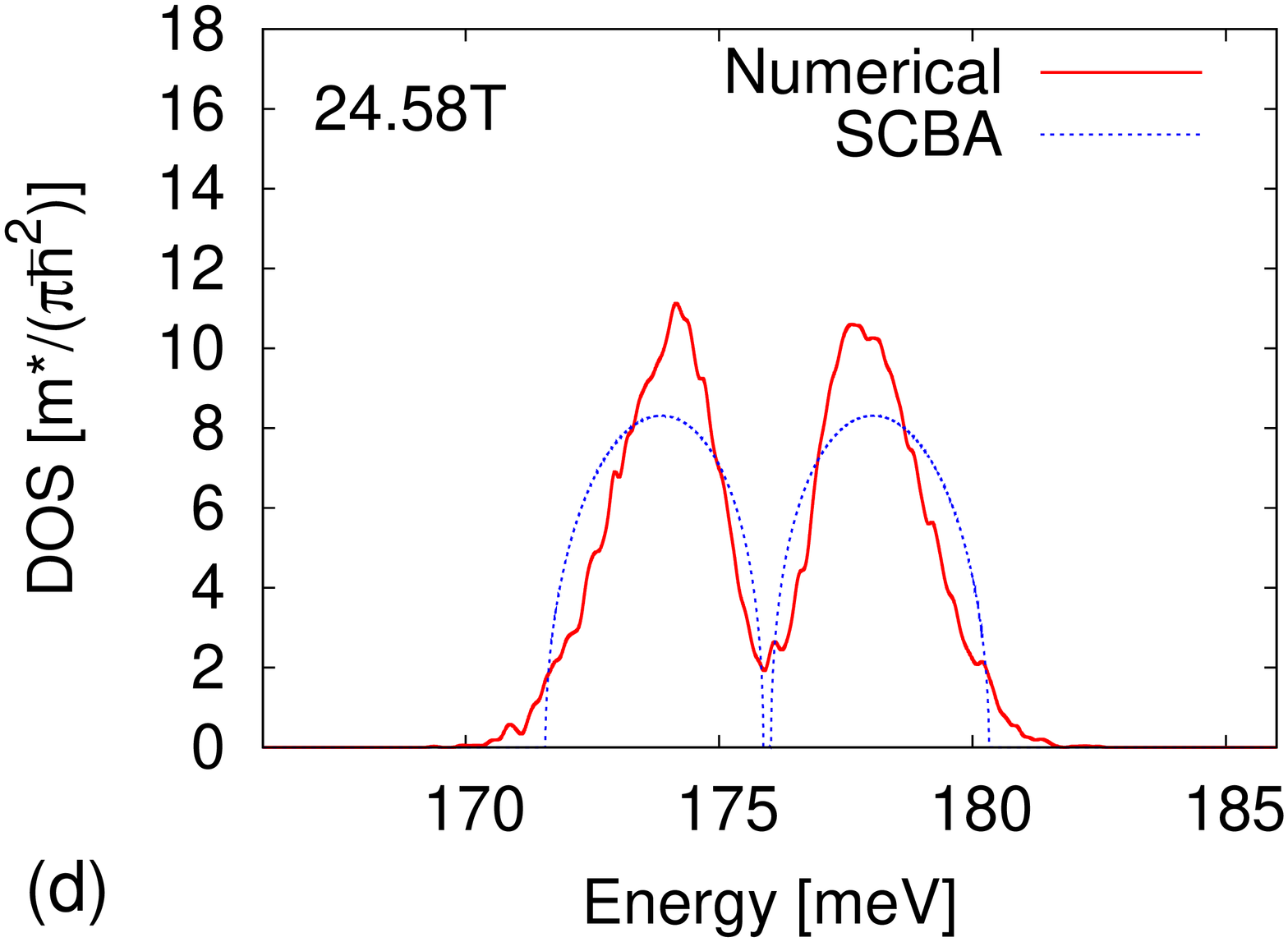}
  \caption{(Color online) Alloy disorder broadening of magneto-polarons: (a) SCBA calculation of the polaron DOS for ${p=1}$ to $4$ at the resonant fields ${B_p}$;
  (b) SCBA calculation of the polaron DOS (in unit of ${m^*/(\pi\hbar^2)}$) around ${B_2=24.58T}$;
  (c) and (d) are numerical results at 23.50T and 24.58T, respectively, to compare with the SCBA calculations.}
\end{figure}

The alloy disorder breaks the ${k_y}$-related translational invariance
 and thus either broadens the magneto-polaron states (weak disorder) or
even destroys the polaron picture (strong disorder).
As for quantitative estimate, by using the self-consistent Born
approximation (SCBA)~\cite{SJ:1999}, we solve a pair of coupled equations of the
self-energy of polarons given by

\begin{eqnarray}
\Sigma_{+p}(\varepsilon)&=&\frac{\overline{V_{+p,+p}^2}}{\varepsilon-\varepsilon_{+p}
-\Sigma_{+p}(\varepsilon)}+\frac{\overline{V_{+p,-p}^2}}{\varepsilon-\varepsilon_{-p}-\Sigma_{-p}(\varepsilon)},{}
\nonumber\\
{}
\\
\Sigma_{-p}(\varepsilon)&=&\frac{\overline{V_{+p,-p}^2}}{\varepsilon-\varepsilon_{+p}
-\Sigma_{+p}(\varepsilon)}+\frac{\overline{V_{-p,-p}^2}}{\varepsilon-\varepsilon_{-p}-\Sigma_{-p}(\varepsilon)},{}
\nonumber\\
{}
\end{eqnarray}
where the alloy squared scattering matrix element is averaged over the alloy
fluctuations~\cite{BrownAustin:2000}, viz.
\begin{equation}
\overline{V_{a,b}^2}=\Big<\sum_{k'_{y}}\Big|\big<a,k_y\big|V_{alloy}\big|b,k^{'}_{y}\big>\Big|^2\Big>_{alloy}.
\end{equation}
For infinitely large (2-dimensional) samples we have
\begin{eqnarray}
\overline{V_{\pm p,\pm p}^2}&=&\frac{D\omega_0}{\Omega_0}\frac{\Big(\frac{3}{2}+\frac{W_p}{\hbar^2\omega^2_{\mp p}}+\frac{U_p}{\hbar^4\omega^4_{\mp p}}\Big)}{\big(1+\Omega^2_{p}/\omega^2_{\mp p}\big)^2}V_{\rm{x}}^2{},
\\
\overline{V_{+p,-p}^2}&=&\frac{D\omega_0}{\Omega_0}\frac{\Big(\frac{3}{2}+\frac{W_p}{\hbar^2\omega_{+ p}\omega_{- p}}+\frac{U_p}{\hbar^4\omega^2_{+ p}\omega^2_{- p}}\Big)}
{\big(1+\Omega^2_{p}/\omega^2_{+ p}\big)\big(1+\Omega^2_{p}/\omega^2_{- p}\big)}V_{\rm{x}}^2{},
\end{eqnarray}
where ${V_{\rm{x}}^2=\rm{x}(1-\rm{x})\big(\Delta
\it{V}\big)^{\rm{2}}}$ and each of the alloy squared scattering matrix elements involves two
alloy scattering events and consequently contains three terms:
zero-phonon term when both events occur among zero-phonon states;
one-phonon term when one event occurs among zero-phonon states and the other among one-phonon states;
two-phonon term when both events occur among one-phonon states.
The one-phonon term has a common factor as
\begin{equation}
W_p=\frac{g^2}{2^{p-1}p!}\sum_{\vec q = ({\vec
q_{/\!\!/},q_z})}\frac{\big|I_p(\vec q)\big|^2}{q^2}\big(\lambda
q_{/\!\!/}\big)^{2p}exp\big(-\frac{1}{2}\lambda^2q^2_{/\!\!/}\big),
\end{equation}
and the two-phonon term's factor reads
\begin{eqnarray}
U_p&=&\frac{3g^4}{2}\sum_{\vec q\vec q'}
\Bigg[\Bigg(\frac{\big|I_{p}(\vec q)I_{p}(\vec
q')\big|}{qq'}\sum_{k=0}^{p}\frac{p!\big(-\lambda^2\Delta
q^{2}_{/\!\!/}\big)^{k}}{(p-k)!(k!)^{2}2^{k}}\Bigg)^{2}{}
\nonumber\\
&&\times exp\big(-\frac{1}{2}\lambda^2\Delta q^2_{/\!\!/}\big)\Bigg].{}
\end{eqnarray}

Once the magneto-polaron self energy is obtained by iteration, the
polaron DOS near ${B_p}$ is given by
\begin{equation}
\rho(\varepsilon)=-\frac{D}{\pi}Im{\left(\frac{1}{\varepsilon-\varepsilon_{+ p}-\Sigma_{+ p}(\varepsilon)}+\frac{1}{\varepsilon-\varepsilon_{- p}-\Sigma_{- p}(\varepsilon)}\right)}.
\end{equation}

As shown in Fig. 5(a), the alloy broadening effect is enhanced with
increasing magnetic field. For larger $p$ the alloy broadening may
be smaller than the polaron splitting. However, when $p$ is large
enough ${B_p}$ is very close to ${B_{p+1}}$ so that the two-LL's
formalism [Eqs. (10) and (11)] is no longer valid. To set a
criterion, we find that a finite polaron gap will open when
${\frac{2\overline{V_{+ p,+ p}^2}}{\sqrt{\overline{V_{+ p,+
p}^2}+\overline{V_{+ p,- p}^2}}}<\hbar\Omega_{p}}$. We can see from
Fig. 5(a) that this criterion is not fulfilled for ${p=1}$ and only
marginally for ${p=2}$, but holds for larger $p$ values with
recognizable double maxima in the DOS. In Fig. 5(b-d) we show more results
for the SCBA polaron DOS of ${p=2}$.

We have also performed a numerical diagonalization of the alloy Hamiltonian
for a given realization of disorder within a basis of polaron
${\big|\pm,k_y\big>}$ states. 
Once the polaron levels are calculated
the density of states is calculated and an average over N=100
samples is performed. As seen in Figs. 5(c) and 5(d), the numerical
calculations agree quite well with the SCBA results though the
broadening effects appear larger in the numerical outputs. We can
see in both SCBA and numerical calculations that out of resonance
(Fig. 5(c)) the DOS displays two peaks of uneven heights.

The DOS peaks of the two broadened LL's appear very asymmetrical at
large detunings. The polaron level that resembles the
${\big|E_{2},0\big>\otimes\big|0_{LO}\big>}$ LL (far from resonance)
acquires a width and a shape that goes smoothly to those of the pure electron
${\big|E_{2},0\big>}$ LL~\cite{BrownAustin:2000}. On the other hand,
at large detuning the SCBA broadening becomes insufficient to fully
account for the alloy broadening. This is because a large number of
uncoupled states become nearly degenerate with the polaron levels
that resemble the one-LO-phonon replica states
${\big|E_{1},p\big>\otimes\big|1_{LO}\big>}$ and these uncoupled
states are not accounted for neither in the SCBA equations nor in
the numerical results in Fig. 5 (see Sec. IV).


\subsection{Time-dependent survival probability\\ in the upper LL's}

\begin{figure}
  \includegraphics[height=.375\textheight,angle=270]{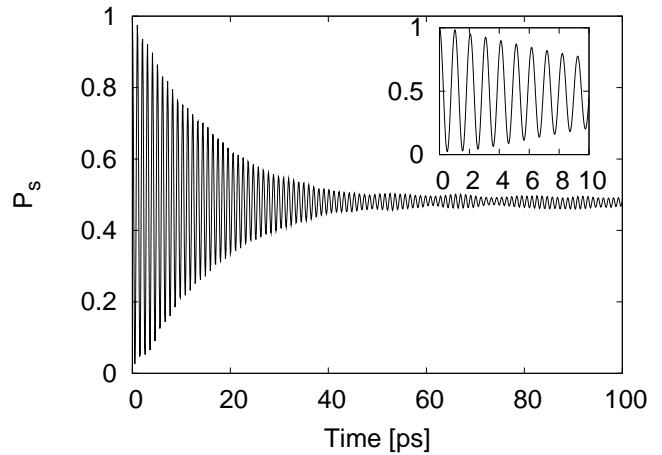}
  \caption{Survival probability $P_s$ for ${B_{2}=24.58T}$ with the initial state as the central level of the broadened ${\big|E_2,0\big>}$ LL's while neglecting all the uncoupled states.}
\end{figure}

The irreversible emission of LO phonon is the most efficient energy
relaxation mechanism in QCL at zero magnetic field provided it is
energy allowed. In the presence of a strong field, in an ideal
sample, this irreversible emission is impossible because of the
formation of intersubband magneto-polarons. Alloy scattering blurs
the magneto-polaron states and it is interesting to ascertain the
nature of the phonon emission in QCL subjected to a strong quantizing magnetic
field. To this end, in the case of the ${p=2}$ resonance, we compute
for a given realization of the alloy disorder the survival
probability in the ensemble of independently broadened LL states
${\big|\nu\big>}$ related to ${\big|E_2,0\big>}$ once the carrier
has been injected in one particular state ${\big|\nu_0\big>}$ of
this ensemble:
\begin{equation}
P_{s}(t)=\sum_{\nu}\Big|\big<\nu \big|\otimes \big<0_{LO}\big|exp\left(\frac{-iHt}{\hbar}\right)\big|\nu_{0}\big >\otimes \big|0_{LO}\big>\Big|^{2},
\end{equation}
where ${H=H_{e}+V_{alloy}+H_{ph}+H_{e-ph}}$. In the following, we
take ${\big|\nu_0\big>}$ as the central level of the broadened
ensemble. Once this is achieved we average over the ${\rm{N}=100}$ realizations of the
alloy disorder. Several magnetic fields have been considered.

When there is a large energy detuning between the zero-LO-phonon
replica of ${\big|E_2,0\big>}$
LL's and the one-LO-phonon replica of
${\big|E_1,p=2\big>}$ LL's, the survival probability very
quickly oscillates between 1 (t = 0) and ${P_{min}}$ to stabilize to
${P_{\infty}=(1+P_{min})/2}$ in a characteristic time ${\tau}$ of a
few picoseconds: if ${\big|\nu_0\big>}$ is the central state of the broadened
LL's and for ${B=20T}$, we find ${\tau\approx 2.8ps}$ and
${P_{\infty}\approx 0.83}$. Right at resonance ${B_2=24.58T}$, see
Fig. 6, the averaged survival probability displays a large number of
fast oscillations (pseudo period: ${1.03ps}$) with a decaying amplitude
(${\tau\approx 42ps}$) that brings $P_s$ to the limiting value
${P_{\infty}=1/2}$.  We note that the pseudo period is close to
${{\pi}/{\Omega_2}=1.04ps}$. The limiting value ${1/2}$ recalls the
fact that the Fr\"ohlich  interaction being so efficient that the carrier
very quickly exits from the initial state. However, once in the
one-LO-phonon replica of the ${p=2}$ LL, there is no sink mechanism the
electron can use to escape from this LL. Thus, another Fr\"ohlich
interaction brings it back into the ${\big|E_2,0\big>}$ LL. There is
no reason a priori for this state to be the same as the initial
state. Subsequently, the electron leaves the initial LL and
oscillates back and forth between the zero-LO-phonon replica
of ${\big|E_2,0\big>}$ LL's and the
one-LO-phonon replica of ${\big|E_1,p=2\big>}$ LL's. This
oscillatory cycle between the two broadened LL's is however
irreversible to the extent that the initial state has very little
chance to be recovered after ${2j}$ jumps. This is in striking contrast
with the coherent polaron oscillations that would result if there
were no alloy scattering acting to blur the polaron states, thereby
leading to an oscillatory cycle between only two polaron states.

The consequence on our understanding of the experimentally observed QCL
oscillatory output versus ${B}$ is significant. Unless one can find a
plausible and fast escape mechanism from the one-LO-phonon replica,
there is no reason to invoke an irreversible escape from the upper
state of the lasing
transition in a QCL. 
The long lived oscillations displayed in Fig. 6 will actually be
limited by the finite lifetime of the phonons due to anharmonicity
effects. This lifetime is about ${10 ps}$ at low temperature and a few
picoseconds at room temperature in bulk GaAs. Similar values were found in
InAs self organized QD's~\cite{Zibik}. Beyond this lifetime,
on average, the oscillations should stop and the electron should
have relaxed to one of the disorder broadened LL states
${\big|E_1,p\big>}$ with no phonon. Note that such a relaxation path
would be in complete contrast with bulk and QW materials at zero
magnetic field but much in agreement with the energy relaxation
scenario established in QD's~\cite{Grange}.

\section{Effects of the uncoupled one-phonon states in the strong coupling regime}

The approximation of neglecting the uncoupled one-phonon states
turns out unsatisfactory because of the very efficient alloy
scattering between magneto-polarons and uncoupled one-phonon states.
As a result, the Rabi oscillations due to polaron effects will be
severely damped since the alloy scattering between polaron states
and the large number of uncoupled one-phonon states will give fast
additional leakage channels for the LL electrons. The uncoupled states
related to the ${\big|\pm p, k_y\big>}$ polaron are approximately
\begin{eqnarray}
\big|uncoupled\big>&=&\big|E_1,p,k_{y}^{'}-q_y\big>\otimes\big|1_{\vec
q}\big>, k_{y}^{'}\neq k_y,{}
\nonumber\\
{}\varepsilon_{uncoupled}&=&\varepsilon_{1,p}+\hbar\omega_{LO}.
\end{eqnarray}
There are ${N_{uncoupled}=([D]-1)N_{phonon}}$ such uncoupled
states for each pair of ${\big|\pm p, k_y\big>}$ (where
${N_{phonon}\geq 480}$ discretized LO phonon modes should be included
to reproduce accurately the polaron gap, and an accurate numerical
diagonalization would give the number in total as
${N_{uncoupled}=[D](N_{phonon}-1)}$). Note also that
these states are energetically placed between the two polaron levels
${\big|\pm p, k_y\big>}$: at their mid-distance at resonance
${B=B_p}$ and tend towards the fast increasing (with field) polaron
branch at high detuning ${\big|B-B_p\big|}$.

The effect of alloy disorder is twofold: (i) broadening the
uncoupled level and (ii) admixing the uncoupled with the polaron
states. However, a full diagonalization, including all polaron and
uncoupled one-phonon states with a large number of phonon modes, is
too heavy numerically (this would mean diagonalizing for each run of
the ${\rm{N}=100}$ a matrix of the size as
${{\rm{Dim}}(H)=\left([D](N_{phonon}+1)\right)^2\ge (2.8\times
10^4)^2}$). In the following, we present estimates illustrating the
importance of effects (i) and (ii).

An estimate of the effect is obtained by remarking that at the
lowest order in ${V_{alloy}}$ it produces a shift of the polaron
states which reads \label{correct}
\begin{equation}
s_{\pm p}=\frac{\big<\sum_{{uncoupled}}{\big|\big<\pm p\big|V_{alloy}\big|uncoupled\big>\big|^2\big>_{alloy}}}{\varepsilon_{\pm p}-\varepsilon_{uncoupled}}.
\end{equation}
It is difficult to give a simplified analytical expression of
${s_{\pm p}}$ like Eqs. (13-14) due to the large value of ${N_{phonon}}$
required. With the results of the numerical diagonalization
as shown in Fig. 5(d), this shift is roughly ${0.4meV}$ at resonance ${B=B_2}$, and
the effect (ii) becomes very important at increasing detuning
${\big|\hbar\delta_2\big|}$ to the polaron branch whose
${\varepsilon_{polaron}}$ tends towards ${\varepsilon_{uncoupled}}$.
For increasing ${\big|\hbar\delta_2\big|}$ it becomes
necessary to take into account higher orders' effects of ${V_{alloy}}$,
once the uncoupled one-phonon states enter the DOS of the closer
broadened polaron branch. Thus it turns out to be extremely inefficient
to handle the problem by making corrections to the SCBA Eqs. (10-11).

\begin{figure}
  \includegraphics[height=.35\textheight,angle=270]{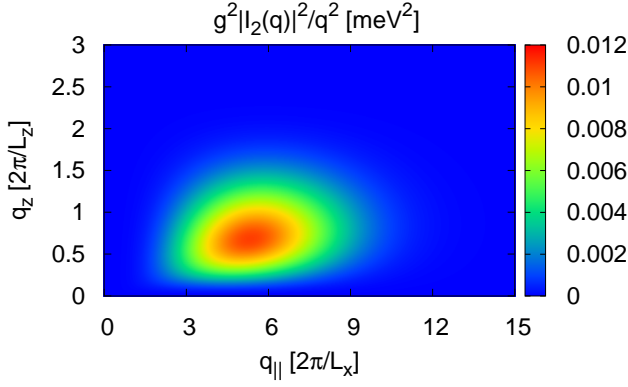}
  \caption{(Color online) Electron-LO-phonon coupling element related to Fr\"ohlich factor versus phonon wave vector.}
\end{figure}

Another way to evidence the importance of the coupling between the polaron and the
unadmixed states consists in looking at the irreversible escape from
one of the polaron states to the unadmixed states broadened by alloy
scattering. An estimate using Fermi's golden rule gives at resonance ${B=B_p}$
\begin{eqnarray}
\frac {\hbar}{2\pi\tau_{\pm p}}&=&\sum_{{\vec q, k_{y}^{'}}}\Big[\big|\beta_{\pm p,\vec q}\big<E_1,p,k_y-q_y\big|V_{alloy}\big|E_1,p,k_{y}^{'}\big>\big|^2
\nonumber\\
&&\times \rho(\varepsilon=\varepsilon_{\pm p})\Big],
\end{eqnarray}
where the DOS for one of the unadmixed states is written
\begin{equation}
\rho(\varepsilon)=\frac {L_{x}L_{y}}{2\pi^2\lambda^2 V_{1p,1p}^{2}}\sqrt{V_{1p,1p}^{2}-\left(\frac {\varepsilon-\varepsilon_{uncoupled}}{2}\right)^{2}} .
\end{equation}
Putting numbers we find ${\tau_{\pm 2}\approx 0.8ps}$, ${i.e.}$
comparable to the polaron period. Thus, it is invalid to neglect the
uncoupled one phonon states and effect (ii) is strong enough to generate
states without well-defined polaronic character.
Note that this result is peculiar to the alloy
scattering we have used for static scatterers. In GaAs based QCL's
the alloy scattering is negligible and the interface defects are
definitely weaker scatterers~\cite{Brown2000,Leuliet}. Hence, for those materials the polaron
levels should be significantly more long-lived.


\begin{figure}
  \includegraphics[height=.317\textheight,angle=270]{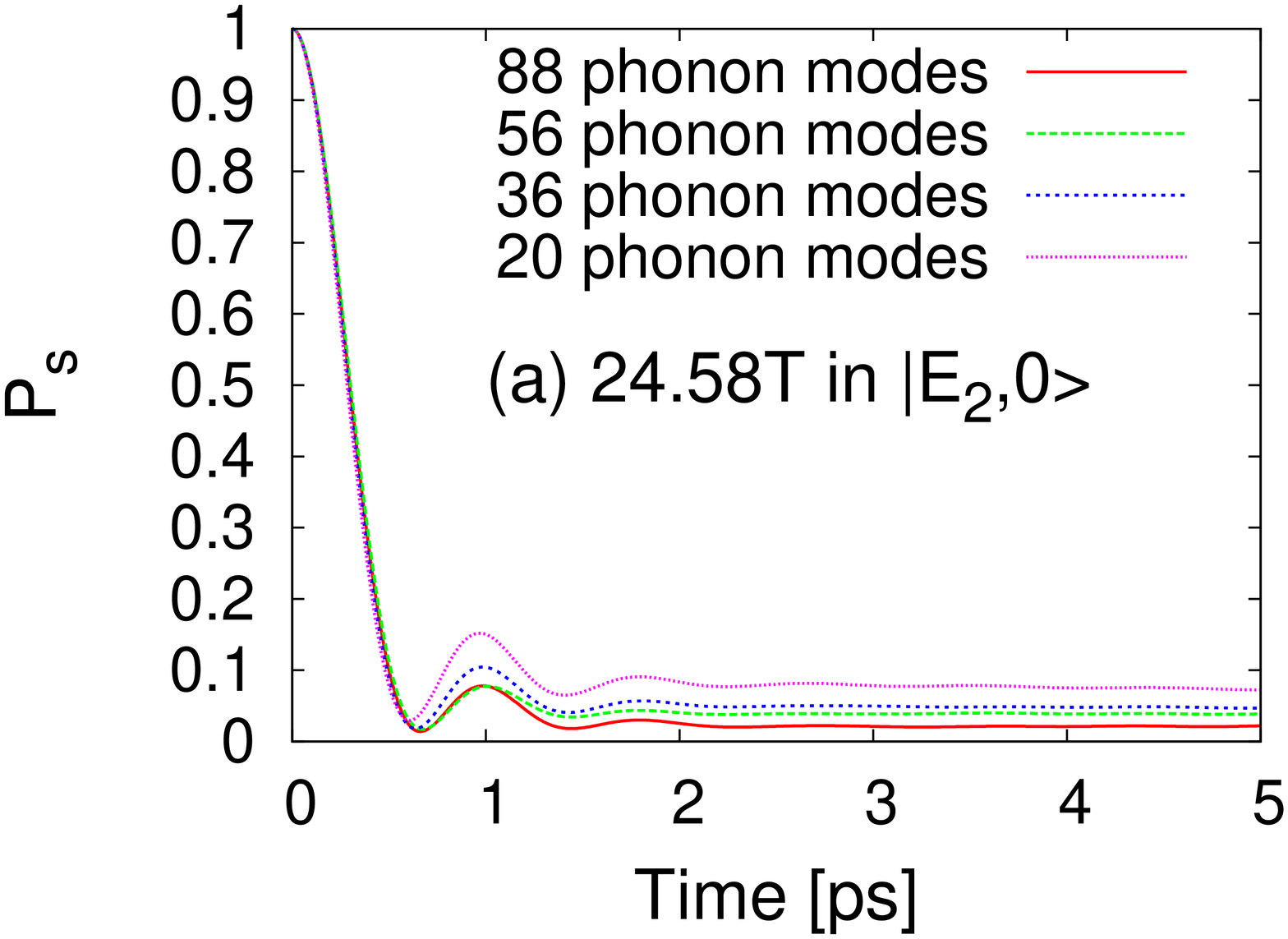}
  \includegraphics[height=.317\textheight,angle=270]{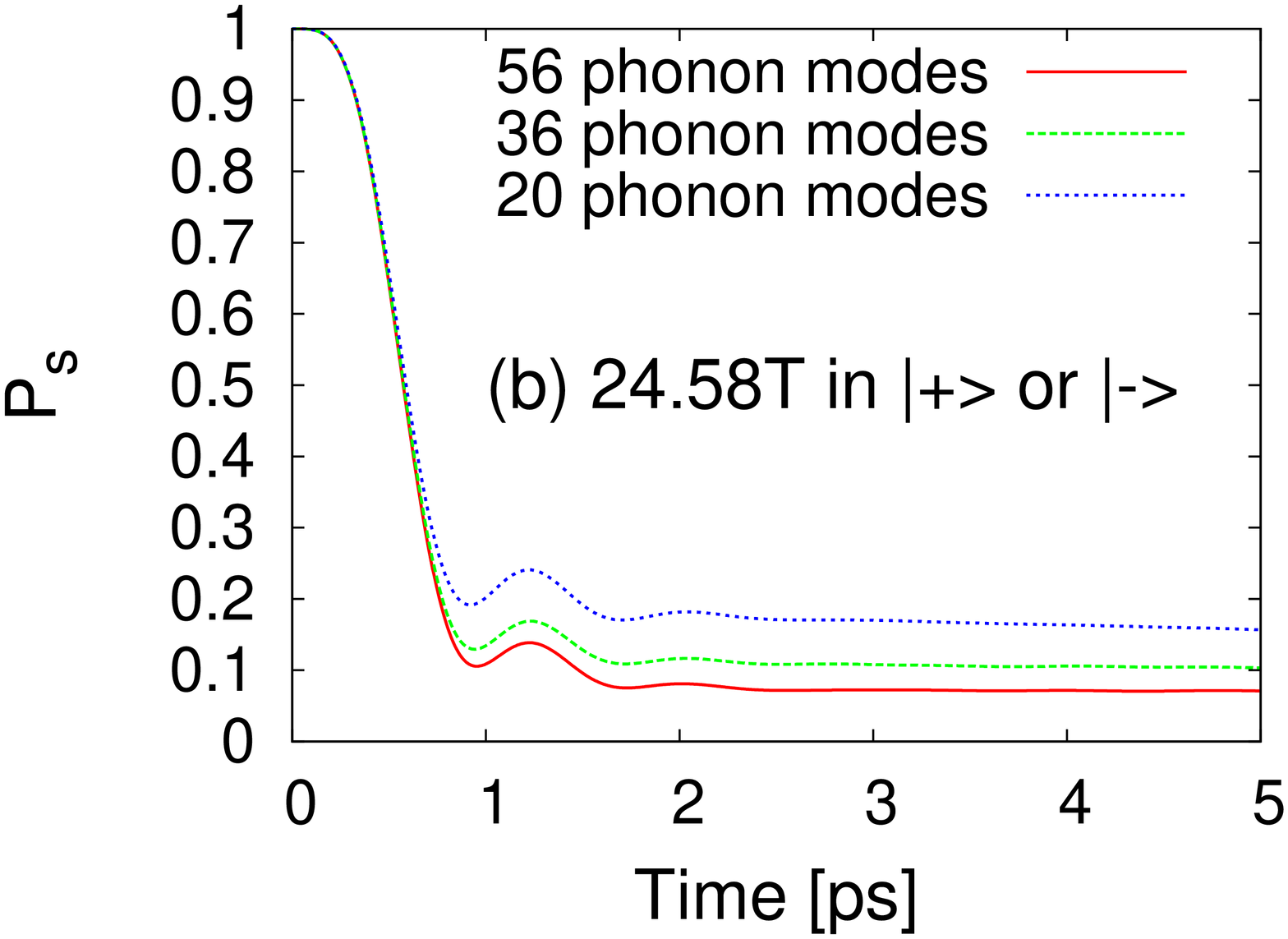}
  \includegraphics[height=.317\textheight,angle=270]{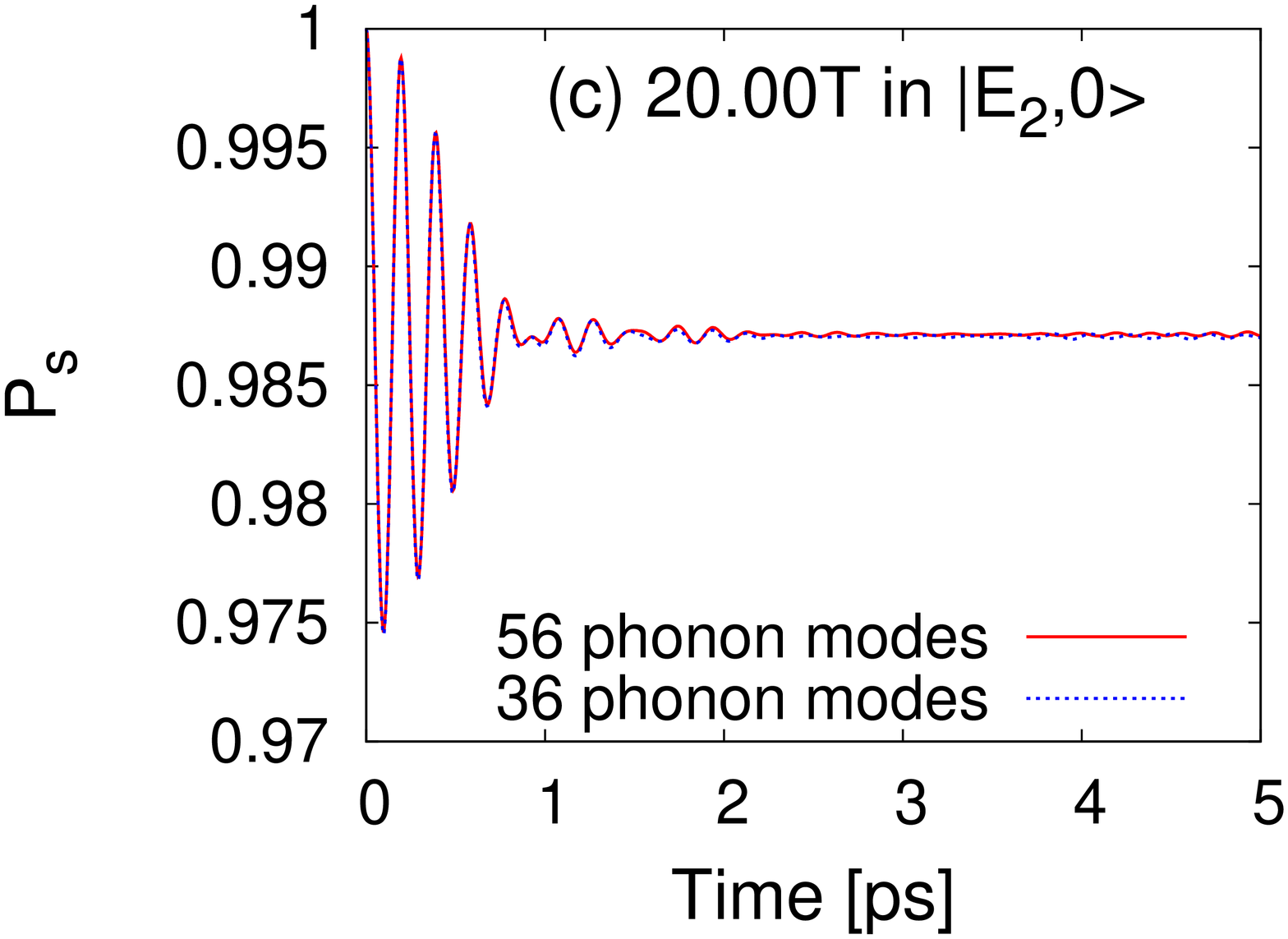}
  \includegraphics[height=.317\textheight,angle=270]{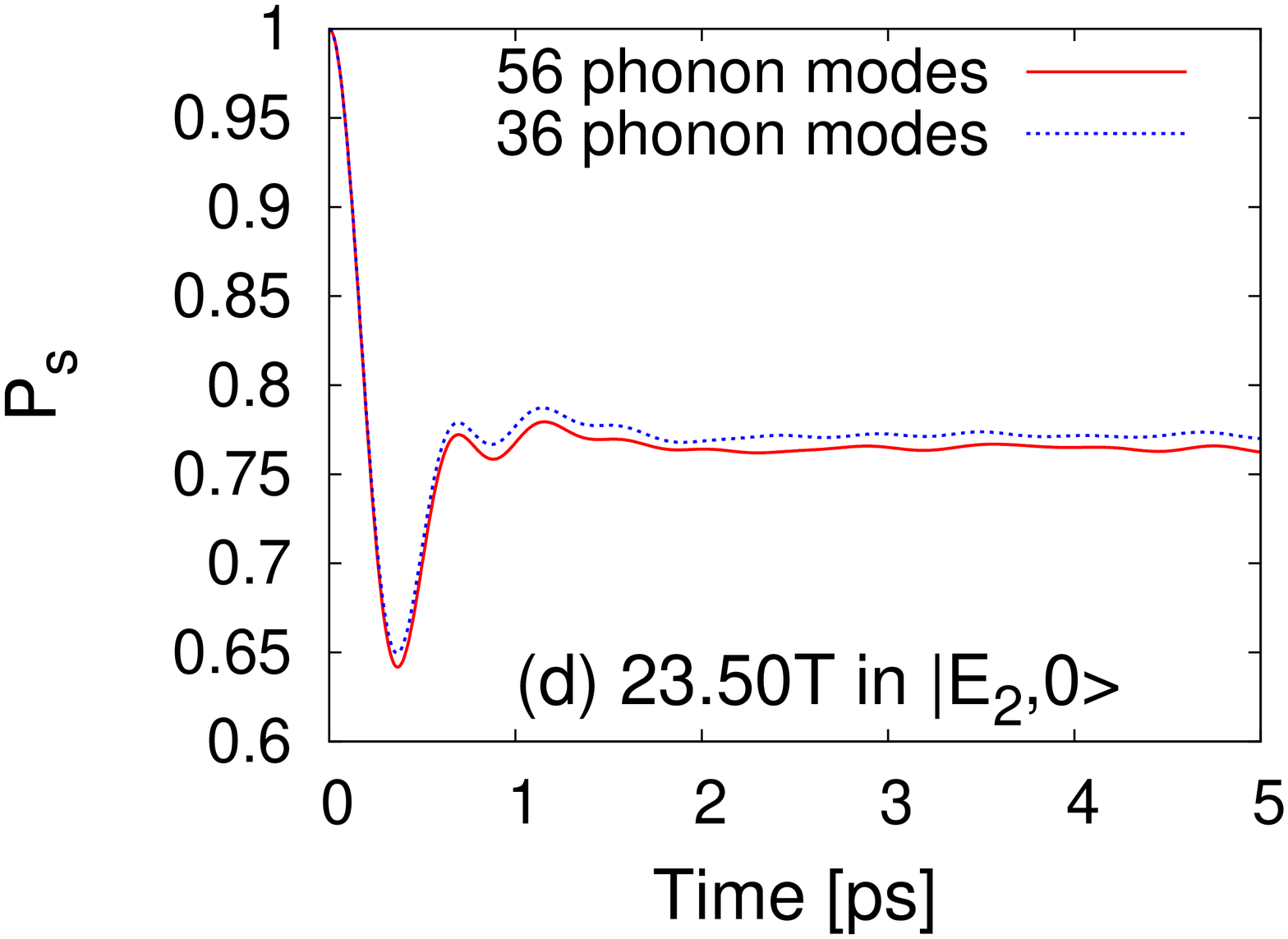}
  \caption{(Color online) Taking increasing numbers of the uncoupled one-phonon states into account, computations of the electron survival probabilities with initial state in center of the alloy broadened LL's ${\big|E_2,0\big>}$: (a) remaining in ${\big|E_2,0\big>}$ at ${B_2=24.58T}$; (b) in polaron states ${\big|\pm\big>}$ at ${B_2}$; (c) in ${\big|E_2,0\big>}$ at 20.00T; (d) in ${\big|E_2,0\big>}$ at 23.50T (see text).}
\end{figure}



Therefore, we have included the uncoupled states in the numerical computation
of the electron time-dependent survival probability in the upper
LL's ${\big|E_2,0\big>}$ (Eq. (18)). We take a certain number of
phonon modes and ensure that the half polaron splitting energy
${\hbar\Omega_2\geq 1.8meV}$ since a small portion of the phonons
give dominant contributions to the electron-phonon coupling (Fig. 7).
As an approximation, we choose a smaller quantization size in the
quantum well free dimension (viz. ${\alpha L_x\times  \alpha
L_y\times L_z}$) for the phonon modes than for the LL's electrons
(which is ${L_x\times L_y\times L_z}$). In the ${z}$-dimension, a
specified wavevector is chosen for the phonon modes
(typically ${2\pi/L_z}$). Numerical
computations of the upper LL's survival probabilities with initially
putting an electron in the center of the LL's ${\big|E_2,0\big>}$
have been done with ${\alpha=1/5}$ (${N_{phonon}=20}$,
${{\rm{Dim}}(H)\approx (1.2\times 10^3})^2$), ${\alpha=1/4}$ (${N_{phonon}=36}$,
${{\rm{Dim}}(H)\approx (2.2\times 10^3)^2}$), ${\alpha=1/3}$ (${N_{phonon}=56}$,
${{\rm{Dim}}(H)\approx (3.4\times 10^3)^2}$) and ${\alpha=1/2}$ (${N_{phonon}=88}$,
${{\rm{Dim}}(H)\approx (5.3\times 10^3)^2}$) (Fig. 8(a))
for ${B_2=24.58T}$ (for ${\alpha=1}$ more than 384 modes are required).
We see that at resonance (Fig. 8(a)) the survival probability in the ${\big|E_2,0\big>}$ LL's
decays very rapidly with time: in about ${0.6ps}$ it has dropped to nearly zero.
The decay is almost insensitive to the number of uncoupled phonon modes and
in fact coincides with the one shown in Fig. 6. However, instead of
recovering to ${1/2}$ at long time, ${P_s}$ remains very small and displays faint remnants
of the polaron oscillations. This is the irreversible oscillation between
the polaron branches and the ${N_{uncoupled}}$ uncoupled phonon modes that
explains the lack of recovering after the first oscillation.

In Fig. 8(b) we show the survival probability of
remaining in the polaron states with the same initial state as that of Fig.
8(a). It should be noticed that at the first half picosecond the
polaron effect is still strong enough to hold its pseudo oscillation
against the damping due to the alloy scattering effect. The same
calculation of the probabilities in ${\big|E_2,0\big>}$ for lower
magnetic fields near ${B_2}$ are shown in Figs. 8(c) and 8(d).
The Figs. reveal that, out of the resonant field and in the
presence of both alloy and LO phonon scatterings, the
probability of relaxing to the lower LL's remains
small at long time.



\begin{figure}
  \includegraphics[height=.38\textheight,angle=270]{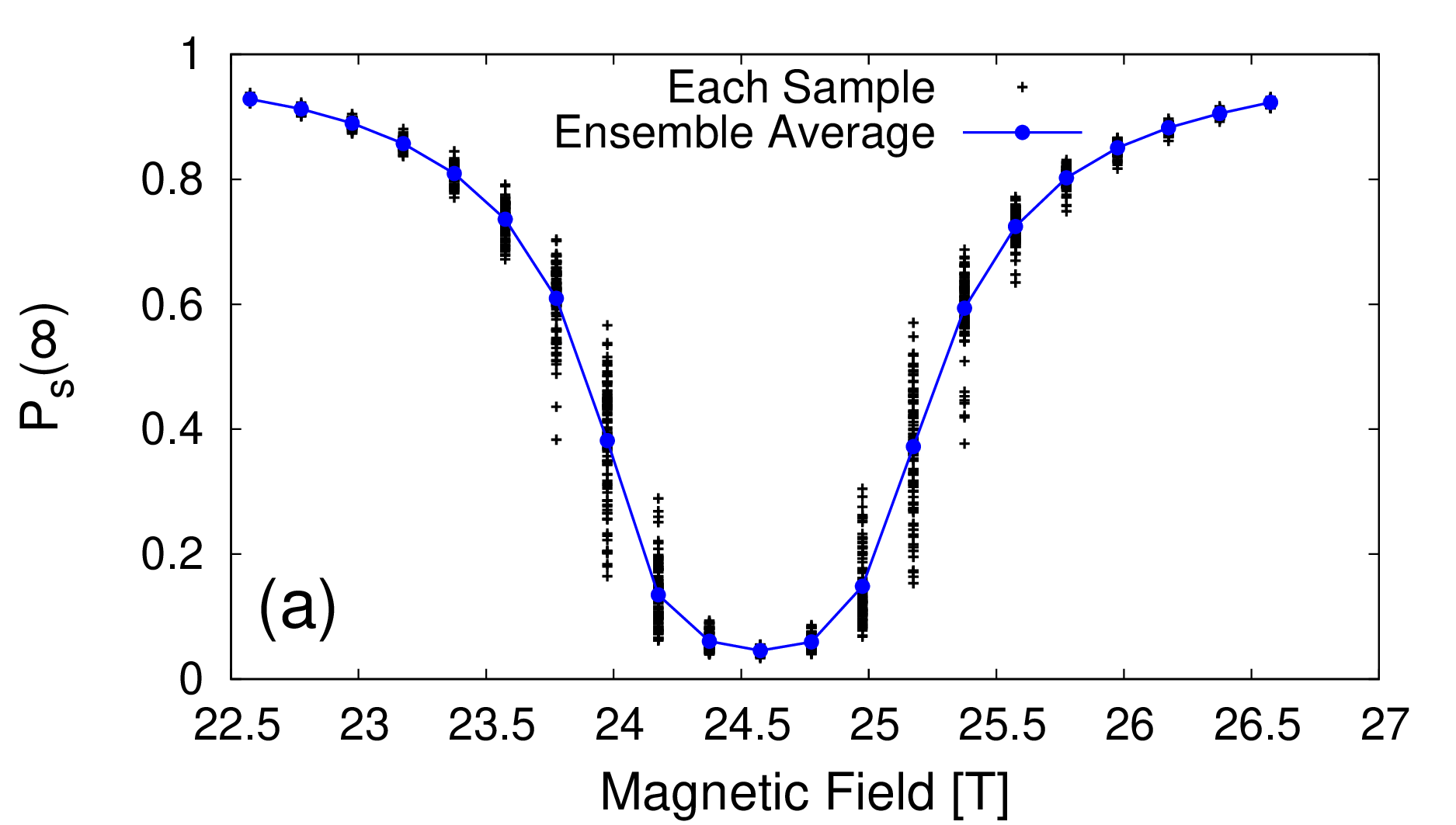}
  \includegraphics[height=.38\textheight,angle=270]{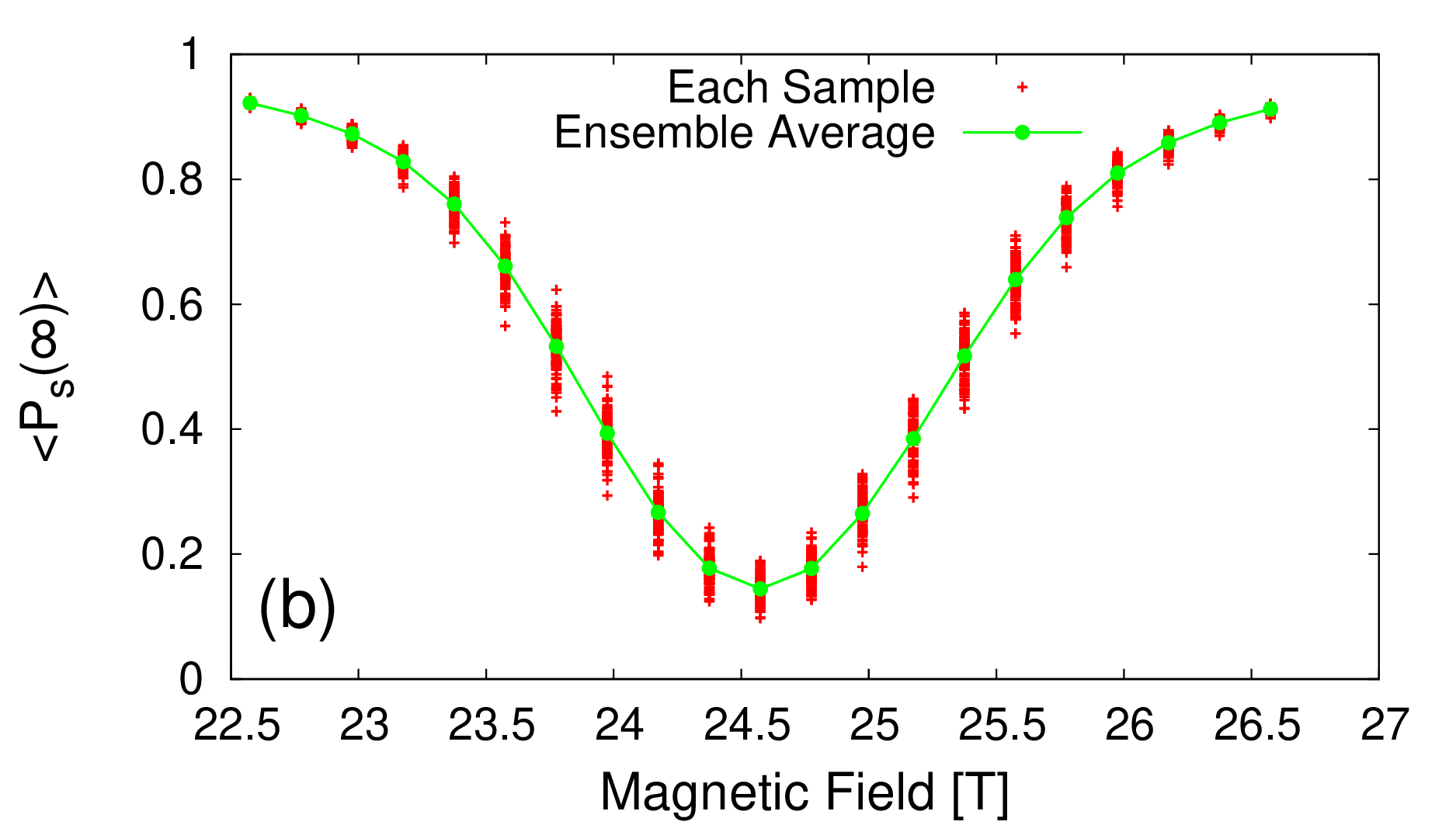}
  \caption{(Color online) Computations with 36 phonon modes of the long time limit of the survival
probability ${P_s(\infty)}$ in the ${\big|E_2,0\big>}$ LL's: (a) Initial states in center of LL's ${\big|E_2,0\big>\otimes\big|0_{LO}\big>}$ for each sample; (b) Average over all initial LL's ${\big|E_2,0\big>\otimes\big|0_{LO}\big>}$ for each sample. Also the ensemble averages are taken over the N=100 samples for each case (jointed by lines, which are guides for the eyes).}
\end{figure}

The experimental implication of our findings on the resonant non-radiative
relaxation in the QCL's subjected to a quantizing magnetic field can be
illustrated in the following way. For one realization of disorder and a
given magnetic field B, we compute the long time limit of the survival
probability ${P_s(\infty)}$ in the ${\big|E_2,0\big>}$ Landau level in two
cases. In case (a), the initial state is in the center of the alloy broadened
${\big|E_2,0\big>\otimes\big|0_{LO}\big>}$ LL's. In case (b),
the quantity ${P_s(\infty)}$ is computed for every state
of ${\big|E_2,0\big>\otimes\big|0_{LO}\big>}$ and its
average ${<P_s(\infty)>}$ over all these initial states
is computed. Then another realization of the disorder is created and both
${P_s(\infty)}$ and ${<P_s(\infty)>}$ are calculated again. Fig. 9 shows the
curves ${P_s(\infty)}$ and ${<P_s(\infty)>}$ versus ${B}$. If alloy scattering
and electron-LO-phonon interactions are the dominant non-radiative
losses, the curves shown in Figs. 9(a) and 9(b) can be compared to
the output power of the QCL. It appears that the calculated widths are
comparable to those seen experimentally~\cite{Brown2000,Leuliet}.

\section{Conclusion}

We have presented a theoretical analysis of the electron-LO-phonon
interaction in QCL's structures in the presence of a strong magnetic
field and of static short ranged scatterers. Our objective was to
ascertain the accuracy of the weak coupling regime between electrons
and LO phonons. Our results show that the very notion of a Fermi's
golden rule is highly questionable in these structures because of
the LL singular density of states. Neither the static scatterers are
weak enough nor the electrons and LO phonons form stable
magneto-polarons. As a result the survival probability in the upper
state of the lasing transition never decays exponentially to zero
but displays a number of damped oscillations before stabilizing to
${1/2}$, thereby evidencing that the magneto-polaron levels never
completely empty. Or, the static scatterers are efficient enough to
wash out the polaron oscillations because it couples these polaron
levels more efficiently to the huge reservoir of uncoupled
one-LO-phonon states than to the polaron states. The survival
probability in the upper state of the lasing transition decay to
zero but not at all in an exponential fashion as would result from
the Fermi's golden rule. Instead, it first decays like in the damped
polaron case and once has reached its first minimum practically
never recovers. The complicated time evolution of the survival
probability evidences the need of a more microscopic description to
understand the physics of the non-radiative mechanisms in QCL's. It
also warns against the estimated efficiency of the static scatterers
or phonon scattering when it is based on oversimplified descriptions
of the disorder on the QCL quantum states.

\begin{acknowledgments}
One of us (Y. C.) would like to thank the French ministry of foreign
affairs for financial support. Chen and Zhu would like to also thank
the supports from the NSFC (Grant No.10774086), and the Basic
Research Program of China (Grant 2006CB921500).
\end{acknowledgments}


\end{document}